\documentclass[prb,aps,twocolumn,10pt,superscriptaddress]{revtex4}
\usepackage{amsmath}
\usepackage{mathtools}
\usepackage{amssymb,wasysym,dsfont}
\usepackage{amsthm}
\usepackage{amsfonts}
\usepackage{enumerate}
\usepackage{verbatim}
\usepackage{latexsym}
\usepackage{bm}
\usepackage{graphicx}
\usepackage{ulem} %using \sout from the ulem package
\usepackage{subfigure}
\usepackage{color}
\usepackage[dvipsnames]{xcolor}
\usepackage[colorlinks]{hyperref}% http://ctan.org/pkg/hyperref
\usepackage{tabularx}

\newcommand{\beq}{\begin{equation}}
\newcommand{\eneq}{\end{equation}}
\input{epsf}

\begin{document}

\title{Quantum interference in transport through almost symmetric double quantum dots}
\author{Zeng-Zhao Li} 
\affiliation{Division of Solid State Physics and NanoLund, Lund University, Box 118, S-22100 Lund, Sweden} 
\author{Martin Leijnse}
\affiliation{Division of Solid State Physics and NanoLund, Lund University, Box 118, S-22100 Lund, Sweden} 
%\affiliation{Center for Quantum Devices and Station Q Copenhagen, Niels Bohr Institute, University of Copenhagen, DK-2100 Copenhagen, Denmark} 

\begin{abstract}
We theoretically investigate transport signatures of quantum interference in highly symmetric double quantum dots in a parallel geometry and demonstrate that extremely weak symmetry-breaking effects can have a dramatic influence on the current. Our calculations are based on a master equation where quantum interference enters as non-diagonal elements of the density matrix of the double quantum dots. We also show that many results have a physically intuitive meaning when recasting our equations as Bloch-like equations for a pseudo spin associated with the dot occupation. In the perfectly symmetric configuration with equal tunnel couplings and orbital energies of both dots, there is no unique stationary state density matrix. Interestingly, however, adding arbitrarily small symmetry-breaking terms to the tunnel couplings or orbital energies stabilizes a stationary state either with or without quantum interference, depending on the competition between these two perturbations. The different solutions can correspond to very different current levels. Therefore, if the orbital energies and/or tunnel couplings are controlled by, e.g., electrostatic gating, the double quantum dot can act as an exceptionally sensitive electric switch. 
\end{abstract}

\date{\today}
\pacs{}
\maketitle

\section{Introduction}

Quantum interference is not only conceptually interesting, but also has a wide range of applications in quantum science and technology. Superconducting quantum interference devices are already standard technology, but quantum interference in electronic transport in nanostructures might be beneficial in as diverse applications as transistors~\cite{Stafford2007} and heat engines.~\cite{Vannucci15prb,SierraSanchez16prb,Lambert16Physique,Samuelsson17prl,Marcos18prb} Molecular junctions and artificial double-quantum-dot molecules have been widely investigated experimentally as typical platforms for demonstrating quantum interference of electrons.~\cite{BayerHawrylak01science,Wiel02rmp,Doty09prl,Guedon12nnano,Garner2018,Miao2018,Nilsson2010,Karlstrom2011} 
Theoretical works have proposed that quantum interference, in contrast to dynamical blockade induced by electron-electron correlations,~\cite{Cottet04prl,Cottet04epl,Belzig05prb,Li13srep} 
is responsible for a bunching behavior (super-Poisson noise or Fano factor larger than $1$) of electrons in transport,~\cite{UrbanKonig09prb,Schaller09prb} 
which might be useful for quantum communication with entangled electrons.~\cite{Burkard00prb} Triple quantum dots~\cite{GaudreauHawrylak06prl,ShimHawrylak09prb} or a single dot with three orbitals allows using quantum interference for demonstrating coherent population trapping of electrons.~\cite{Michaelis06epl,XuSham08nphys,Bayer08nphys} The formed trapping state -- or dark state -- is potentially useful for cooling a nanomechanical resonator.~\cite{Li11epl}
% being useful for, e.g.,  quantum information processing~\cite{TianZoller04prl,VitaliAspelmeyer07prl,HartmannPlenio08prl}.

In devices where multiple quantum dots or molecules are parallel-coupled to two leads, the importance of quantum interference depends in a complicated way on the relative strengths and phases of the various different tunnel couplings. The dependence on the energy difference between the dot orbitals seems less complicated, with quantum interference being suppressed whenever this energy difference exceeds the energy scale set by the tunnel coupling. The quasi-degenerate case of orbital splitting and tunnel coupling of comparable magnitude was investigated in Ref.~\onlinecite{SchultzOppen09prb} by deriving a master equation in the singular-coupling limit.~\cite{Breuer02}
Note, however, that a later work, that focused on interference effects in a benzene molecule, showed that the master equation in the presence of quasidegenerate states could also be derived under the widely used weak-coupling approximation.~\cite{DarauDonariniGrifoni09prb}
%
%Regarding the symmetry breaking due to perturbations to tunnelling couplings, an indirect tunnelling coupling parameter was introduced to study effects of quantum interference on transport through parallel double quantum dots both without \cite{Kubo06prb} and with \cite{Tokura07njp} Coulomb interaction, and an interference-induced suppression of current was observed \cite{Tokura07njp}. 
%
%Besides two types of perturbations that break symmetry as mentioned, it should be further noted that the use of a fully numerical simulation to the study a highly symmetric system (e.g., an exact degeneracy) seems impossible due to an invertible matrix involved in solving a matrix equation, %\cite{LiLeijnse18} and thus one has to be much more careful in particular about an emergence of quantum interference from the symmetry breaking in the system.  Therefore, a systematic understanding of how symmetry breaking via small perturbations affects quantum interference in highly symmetric systems becomes important even in a simple system of double quantum dots or molecular junctions. 

Here, we focus on the case of an almost fully symmetric system, including only small perturbations away from orbital degeneracy and tunnel-coupling symmetry. We calculate the stationary state electric current through an interacting double quantum dot parallel-coupled between two leads using a master equation approach. The master equation is solved for the (reduced) density matrix describing the nonequilibrium state of the double dot. To make a clear connection to previous works on transport in quantum dot spin-valves (see, e.g., Refs.~\onlinecite{BraunKonig04prb} and~\onlinecite{Hell15prb}), we recast the master equation as a Bloch-like equation for a pseudo-spin representation of the dot occupation. 

In the fully symmetric case (equal energies of the two quantum dot orbitals and equal tunnel couplings to both dots and both leads), the rate matrix appearing in the master equation becomes singular, signaling that there is no unique stationary state (a similar instability in a different model was studied in Ref.~\onlinecite{Holubec18jltp}). 
This case is, however, not experimentally relevant as energies and couplings will never be exactly equal. We therefore investigate the effects of weak perturbations to these parameters, where weak means much smaller than all other energy scales of the system. We find that the interplay between perturbations to orbital degeneracy and tunnel couplings dramatically affects the signatures of quantum interference in electron transport. In the presence of a small breaking of orbital degeneracy, but fully symmetric tunnel couplings, there is no quantum interference between electrons tunneling through the two dots (the density matrix of the double quantum dot, written in the basis of occupation of the individual dots, is diagonal). The current is independent of the magnitude of the breaking of orbital degeneracy, as long as it remains the smallest energy scale of the problem. In contrast, when the symmetry of the tunnel couplings is weakly broken, but the orbital degeneracy remains exact, the current is significantly suppressed due to destructive interference (non-diagonal elements of the density matrix). Again, the current is independent of the strength and exact details of the symmetry breaking (as long as it remains weak). When both the orbital degeneracy and the tunnel-coupling symmetry are weakly broken, the current becomes highly sensitive to the details and strengths of the different symmetry-breaking terms. In particular, for certain ranges of orbital energies, extremely small variations in the tunnel couplings or orbital energies switch the current between very close to zero and a large value by turning quantum interference on or off. These changes can be induced by gate voltages and such sensitive switching behavior -- not limited by temperature -- is highly desirable in transistors and related applications. We also find strong negative differential resistance and rectifying behavior.  

This paper is organized as follows. In Sec.~\ref{sec:model}, we introduce our double quantum dot model.  
In Sec.~\ref{sec:approach}, a master equation approach and also the formalism of a pseudo spin are presented. In Sec.~\ref{sec:Qinfer}, we demonstrate that there is no unique stationary state solution for the completely symmetric case and explain how the singularity appears in the pseudo-spin formalism. The symmetry-breaking effects due to small orbital detuning and tunnelling deviation on the quantum interference are illustrated in Sec.~\ref{sec:pertur}. Summary and conclusions are finally given in Sec.~\ref{sec:discuss}.

\section{Model \label{sec:model}}

\begin{figure}[t]
\centering
  \includegraphics[width=.85\columnwidth]{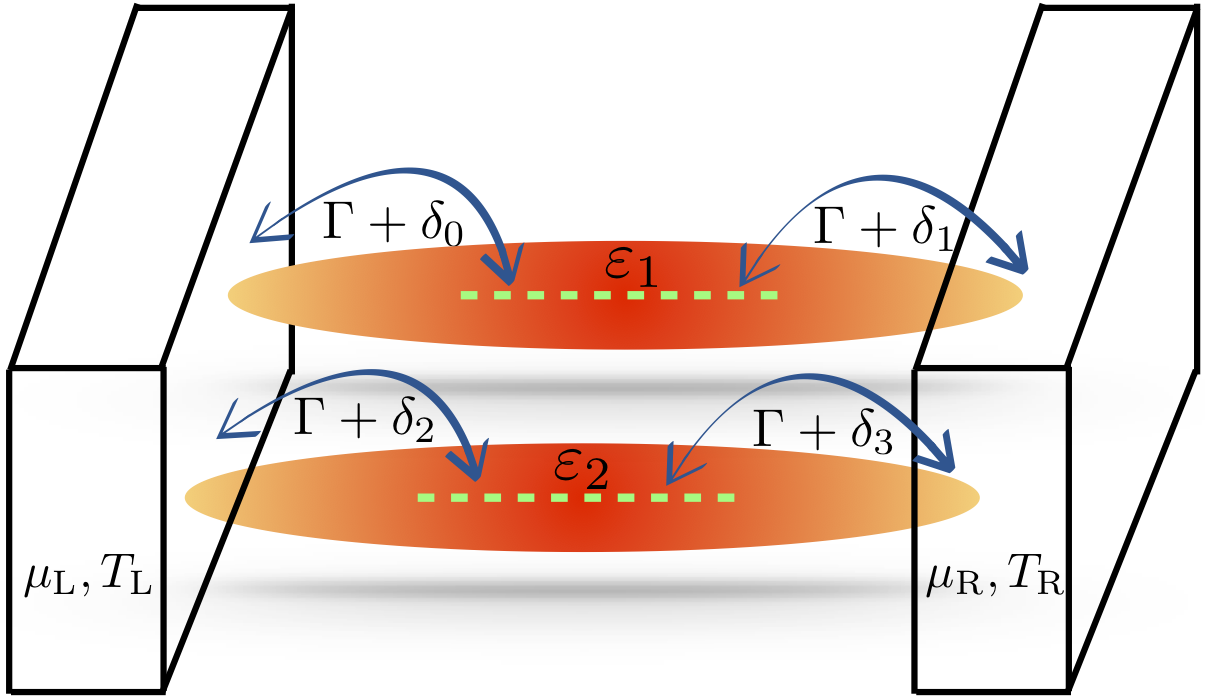} 
\caption{(Color online) Two single-orbital quantum dots coupled in parallel between two leads. The symmetry of the system can be broken by lifting the orbital degeneracy ($\delta\varepsilon = \varepsilon_2 - \varepsilon_1 \neq 0)$ and/or by unequal tunnelling rates ($\delta_j \neq 0$, $j=0,1,2,3$).}
\label{fig:schematic_N2}
\end{figure}

The system of two single-orbital quantum dots in a parallel geometry, coupled to both left and right leads is schematically shown in Fig. \ref{fig:schematic_N2}, and modelled by the Hamiltonian (we use $\hbar = k_B = 1$ throughout the rest of the paper)
\begin{eqnarray}
H&=&H_{\rm S}+H_{\rm B}+H_{\rm T},
\end{eqnarray}
where%with ($\hbar=1$) 
\begin{eqnarray}
H_{\rm S}&=&
\sum_{\substack{i=1,2}} \varepsilon_{i} d_{i}^{\dagger}d_{i} 
+U d_{1}^{\dagger}d_{1} d_{2}^{\dagger}d_{2},  \label{eq:H_S} \\
H_{\rm B}&=&\sum_{\substack{k,s={\rm L,R}}} \omega_{k,s}c_{k,s}^{\dagger}c_{k,s}, \label{eq:H_B}\\
H_{\rm T}&=&\sum_{\substack{i=1,2}}\sum_{k,s={\rm L,R}} 
t_{k,s,i} c_{k,s}^{\dagger}d_i + t_{k,s,i}^\ast{} d_i^{\dagger}c_{k,s}. \label{eq:H_int}
\end{eqnarray}
Here, $H_{\rm S}$ with fermionic operators $d_{i}^{\dagger}$ and $d_{i}$ describes the two single-orbital quantum dots  
with $\varepsilon_{i}$ being the onsite orbital energy of the $i$th dot  
and $U$ the Coulomb interaction between two electrons occupying different quantum dots. For conceptual simplicity we neglect the spin degree of freedom, knowing that the density matrix in the spinful case will remain diagonal in spin for nonmagnetic leads and in the absence of spin-orbit coupling.
The left and right leads are described by $H_{\rm B}$ with fermionic operators $c_{k,s}^{\dagger}$ and $c_{k,s}$. 
In the tunnel Hamiltonian $H_{\rm T}$, $t_{k,s,i}$ is the tunnelling amplitude between the $i$th quantum dot and the $s$th lead.
We emphasize that the double quantum dot considered in our work is a parallel coupled double dot (rather than a serial coupled) and that there is no tunnel coupling between the dots. 

The eigenstates of the isolated double quantum dot are 
$\{|a\rangle, |b\rangle,|c\rangle,|d\rangle\}$  %$\{|00\rangle, |10\rangle,|01\rangle,|11\rangle\}$ 
with $|a\rangle=|00\rangle$, $|b\rangle=|10\rangle$, $|c\rangle=|01\rangle$, and $|d\rangle=|11\rangle$ where $n_1$ and $n_2$ in $|n_1n_2\rangle$ represent the occupation numbers of the 1st and 2nd dots, respectively. The double dot Hamiltonian in this basis becomes
\begin{eqnarray}
H_{\rm S} &=& \varepsilon_1|b\rangle\langle b|+\varepsilon_2|c\rangle\langle c| 
+(\varepsilon_1+\varepsilon_2+U)|d\rangle\langle d| , \label{eq:H_S_eigen} 
\end{eqnarray}
while the tunnel Hamiltonian is 
\begin{eqnarray}
H_{\rm T} &=& \sum_{\substack{k,s={\rm L,R} }} [t_{k,s,1}^{\ast}
(|b\rangle\langle a| +|d\rangle\langle c|) \notag\\ 
&&+t_{k,s,2}^{\ast}(|c\rangle\langle a| -|d\rangle\langle b|) ] c_{k,s} +H.c.
%\notag\\
%&&+t_{k,s}^{\ast} c_{k,s}^{\dagger}
%(|00\rangle\langle 10| +|01\rangle\langle 11| +|00\rangle\langle 01|++|10\rangle\langle11|)
. \label{eq:H_int_eigen}
\end{eqnarray}  

%The highly symmetric or symmetry-preserved system under our following consideration means identical on-site energies i.e., $\varepsilon_i=\varepsilon=-eV_{g}$ ($i=1,2$) and tunnelling coupling [i.e., $\delta\varepsilon=\delta_j=0$ ($j=0,1,2,3$) in Fig.~\ref{fig:schematic_N2} and the relation between the coupling strength e.g., $t_{k,s,i}$ in Eq.~(\ref{eq:H_int_eigen}) and tunnelling rate $\Gamma$ is given below]. 

\section{Master equation \label{sec:approach}}

Following the standard procedures  such as the second-order perturbation theory, the Born-Markovian approximation, and tracing out the degree of freedom of the leads~\cite{Blum96,ScullyZubairy1997,Breuer02}, the evolution of reduced double dot density matrix  
is governed by the equation %(see Appendix~\ref{sec:Append_Deriv}) 
\begin{eqnarray}
\dot{\rho}&=&-i[H_{\rm S},\rho]+\mathcal{L}_{\square}\rho+\mathcal{L}_{\boxtimes}\rho+\mathcal{P}\rho, 
\label{eq:MBME}
\end{eqnarray}
where $H_{\rm S}$ is given in Eq.~(\ref{eq:H_S_eigen}) and the superoperators describing dissipation of the system due to the couplings to the leads are
%\begin{widetext} 
\begin{eqnarray}
\mathcal{L}_{\square}\rho&=&\sum_{s={\rm L,R}} 
\Gamma_{s1} \{ f_s(\varepsilon_1) \mathcal{D}[|b\rangle\langle a|]\rho 
+\bar{f}_s(\varepsilon_1) \mathcal{D}[||a\rangle\langle b|]\rho \notag\\
&&+f_s(\varepsilon_1+U) \mathcal{D}[|d\rangle\langle c|]\rho 
+\bar{f}_s(\varepsilon_1+U) \mathcal{D}[|c\rangle\langle d|]\rho \} \notag\\
&&+\sum_{s={\rm L,R}} \Gamma_{s2} \{ f_s(\varepsilon_2) \mathcal{D}[|c\rangle\langle a|]\rho  
+\bar{f}_s(\varepsilon_2) \mathcal{D}[||a\rangle\langle c|]\rho \notag\\
&& +f_s(\varepsilon_2+U) \mathcal{D}[|d\rangle\langle b|]\rho 
+\bar{f}_s(\varepsilon_2+U) \mathcal{D}[|b\rangle\langle d|]\rho \}, \notag \\ %\\
\label{eq:Lindbladian_noInterfer}
\end{eqnarray} 

\begin{eqnarray}
\mathcal{L}_{\boxtimes}\rho &=& \sum_{s={\rm L,R}}
\sqrt{\Gamma_{s1}\Gamma_{s2}} \{ f_s(\varepsilon_1) \mathcal{D}[ |c\rangle\langle a|, |b\rangle\langle a|] \rho \notag\\
&& +\bar{f}_s(\varepsilon_1) \mathcal{D}[|a\rangle\langle c|,   |a\rangle\langle b|] \rho \notag\\
&&- 
f_s(\varepsilon_1+U) \mathcal{D}[ |d\rangle\langle b|, |d\rangle\langle c|] \rho  \notag\\
&& -\bar{f}_s(\varepsilon_1+U)  \mathcal{D}[|b\rangle\langle d|,  |c\rangle\langle d|]\rho \} \notag\\
&& + \sum_{s={\rm L,R}} \sqrt{\Gamma_{s1}\Gamma_{s2}} 
\{ f_s(\varepsilon_2) \mathcal{D}[|b\rangle\langle a|,  |c\rangle\langle a|] \rho \notag\\
&& + \bar{f}_s(\varepsilon_2) \mathcal{D}[ |a\rangle\langle b|, |a\rangle\langle c|] \rho \notag\\
&&- f_s(\varepsilon_2+U) \mathcal{D}[|d\rangle\langle c|,  |d\rangle\langle b|] \rho  \notag\\
&&  -\bar{f}_s(\varepsilon_2+U) \mathcal{D}[ |c\rangle\langle d|, |b\rangle\langle d|] \rho  \}, 
\label{eq:Lindbladian_yesInterfer}
\end{eqnarray}
and
\begin{eqnarray}
\mathcal{P}\rho &=& \frac{i}{\pi}
\sum_{s={\rm L,R}} \Gamma_{s1} \{ p_s(\varepsilon_1)[|a\rangle\langle a|-|b\rangle\langle b|,\rho] \notag\\
&&+p_s(\varepsilon_1+U)[|c\rangle\langle c|-|d\rangle\langle d|,\rho] \} \notag\\
&& + \Gamma_{s2} \{ p_s(\varepsilon_2)[|a\rangle\langle a|-|c\rangle\langle c|,\rho] \notag\\
&&+p_s(\varepsilon_2+U)[|b\rangle\langle b|-|d\rangle\langle d|,\rho] \} \notag\\
&& +\sum_{s={\rm L,R}} \sqrt{\Gamma_{s1}\Gamma_{s2}} \notag\\
&&\times \{ p_s(\varepsilon_1) ([|a\rangle\langle c|,[|b\rangle\langle a|,\rho]] 
-[|c\rangle\langle a|,[|a\rangle\langle b|,\rho]]) \notag\\
&&-p_s(\varepsilon_1+U) ([|b\rangle\langle d|,[|d\rangle\langle c|,\rho]]-[|d\rangle\langle b|,[|c\rangle\langle d|,\rho]]) \notag\\
&&+p_s(\varepsilon_2) ([|a\rangle\langle b|,[|c\rangle\langle a|,\rho]]-[|b\rangle\langle a|,[|a\rangle\langle c|,\rho]]) \notag\\
&&-p_s(\varepsilon_2+U) ([|c\rangle\langle d|,[|d\rangle\langle b|,\rho]]-[|d\rangle\langle c|,[|b\rangle\langle d|,\rho]]) \}. \notag\\
\label{eq:Lindbladian_principalpart}
\end{eqnarray} 
%\end{widetext}
%\end{widetext}
Explicit expressions for the different matrix elements of $\rho$ are given in Appendix~\ref{sec:Append_Deriv}. For the calculations presented in this work, we have in part relied on the numerical implementation described in Ref.~\onlinecite{qmeq17}.  In Eq.~(\ref{eq:Lindbladian_principalpart}), $p_{s}(\omega) = \Re [\Psi (\frac{1}{2} + \frac{i}{2\pi} \frac{\omega-\mu_s}{T_s})]$, where $\Re[\cdot]$ denotes the real part and the digamma function $\Psi$ originates from principal value integrals. $\mu_s$ and $T_s$ are the chemical potential and temperature of lead $s$. In all calculation we consider a symmetric voltage bias, $\mu_L=-\mu_R=eV_b/2$, and equal temperatures, $T_L = T_R = T$. 
$f_s(\omega)$ in Eqs.~(\ref{eq:Lindbladian_noInterfer}) and (\ref{eq:Lindbladian_yesInterfer}) is the Fermi-Dirac distribution and $\bar{f}_s(\omega) = 1-f_s(\omega)$. We have furthermore defined 
\begin{eqnarray}
\mathcal{D}[A]\rho&=&2A\rho A^{\dagger}-\rho A^{\dagger}A-A^{\dagger}A\rho, \label{eq:superOper_D}\\
\mathcal{D}[ A,B]\rho&=&  A \rho B^{\dagger}+  B\rho A^{\dagger}
- \rho B^{\dagger}A -A^{\dagger} B \rho . \label{eq:superOper_G}
\end{eqnarray}  
Note that $\mathcal{D}[A]\rho=\mathcal{D}[ A,A]\rho$ which only involves a single pathway of electron tunnelings. The bare tunnelling rate is $\Gamma_{si}= \pi |t_{k,s,i}|^2 \varrho_s$ with $\varrho_{\rm L (R)}$ being the density of states in the left (right) lead which is assumed to be constant. We parametrize the tunnel couplings as (see Fig.~\ref{fig:schematic_N2})
\begin{eqnarray}
\Gamma_{L1}&=&\Gamma+\delta_0, \text{  } \Gamma_{R1}=\Gamma+\delta_1, \\
 \Gamma_{L2}&=&\Gamma+\delta_2, \text{  } \Gamma_{R2}=\Gamma+\delta_3,
\end{eqnarray}  
where $\delta_j$ ($j=0,1,2,3$) is a (small) perturbation to the tunnelling rate $\Gamma$. We also let $\delta \varepsilon = \varepsilon_2 - \varepsilon_1$ denote the energy difference between dot orbitals and $\varepsilon = (\varepsilon_1 + \varepsilon_2)/2$ denote their avarage which can be controlled by a gate voltage via $\varepsilon=-e\alpha V_g$, where we set $\alpha=1$ for simplicity. In the fully symmetric case we have $\delta \varepsilon = \delta_j = 0$.  

In Eq.~(\ref{eq:MBME}), the first term describes the free evolution of the double quantum dot. $\mathcal{L}_{\square}\rho$ given by Eq.~(\ref{eq:Lindbladian_noInterfer}) involves the well-known form $\mathcal{D}[A]\rho$ [Eq.~(\ref{eq:superOper_D})] and corresponds to tunnelling processes without interference of electrons. Thus, including only this term is equivalent to the Pauli rate equation for the diagonal stationary density matrix [the first term in Eq.~(\ref{eq:MBME}) vanishes in this case].
In addition, Eq.~(\ref{eq:MBME}) explicitly contains $\mathcal{L}_{\boxtimes}\rho$ [Eq.~(\ref{eq:Lindbladian_yesInterfer})] involving $\mathcal{D}[ A,B]\rho$ ($A\neq B$) that is responsible for quantum coherence between states where an electron occupies dot 1 and dot 2, mediated by the electrodes. Note that $\mathcal{L}_{\boxtimes}\rho$ would not survive the rotating-wave approximation for large $\delta \varepsilon$. 
Finally, Eq.~(\ref{eq:MBME}) includes the term $\mathcal{P}\rho$ [Eq.~(\ref{eq:Lindbladian_principalpart})] which originates from principal value integrals and describes a tunneling-induced shift of the quantum dot orbitals.

%To illustrate the quantum interference in the highly symmetric system and also its response to perturbations that break symmetry, here we consider a steady-state current $I=\langle I_s\rangle$ %=Tr_{\rm S+B}(I_s\rho_t)

The current operator is given by $I_s=ie\sum_{k,s,i} t_{k,s,i} c_{k,s}^{\dagger}d_i  - H.c.$. Here we consider the stationary state current flowing into lead R, $I = \langle I_{\rm R} \rangle$ which is given by 
\begin{eqnarray}
\frac{I}{-e}&=& 2\{
[\Gamma_{{\rm R}1}\bar{f}_{\rm R}(\varepsilon_1+U)+\Gamma_{{\rm R}2}\bar{f}_{\rm R}(\varepsilon_2+U)]  \rho_{dd} \notag\\
&& -[\Gamma_{{\rm R}1} f_{\rm R}(\varepsilon_1)+\Gamma_{{\rm R}2} f_{\rm R}(\varepsilon_2)] \rho_{aa} \notag\\
&& +[\Gamma_{{\rm R}1} \bar{f}_{\rm R}(\varepsilon_1) - \Gamma_{{\rm R}2} f_{\rm R}(\varepsilon_2+U)] \rho_{bb} \notag\\
&& + [\Gamma_{{\rm R}2} \bar{f}_{\rm R}(\varepsilon_2) - \Gamma_{{\rm R}1} f_{\rm R}(\varepsilon_1+U)] \rho_{cc}  \} \notag\\
&& + \sqrt{\Gamma_{{\rm R}1}\Gamma_{{\rm R}2}} [\bar{f}_{\rm R}(\varepsilon_1) + \bar{f}_{\rm R}(\varepsilon_2) \notag\\
&& + f_{\rm R}(\varepsilon_1+U) + f_{\rm R}(\varepsilon_2+U)]  
(\rho_{bc}+\rho_{cb}) ,  \label{eq:current}
\end{eqnarray}
where $\rho_{\alpha\beta}$ ($\alpha,\beta=a,b,c,d$) are the elements of the stationary state density matrix.

In addition to the standard form of the master equation, we consider Bloch-like equations for pseudo-spin components in order to help understand and interpret our results. We define the $z$ component of the pseudo spin as being proportional to the charge difference between the two dots. Similar to the case of a real spin,~\cite{BraunKonig04prb} the relation between pseudo-spin components and reduced density matrix elements is then given by 
\begin{eqnarray}
S_x &=& \frac{\rho_{bc}+\rho_{cb}}{2}, S_y = i\frac{\rho_{bc}-\rho_{cb}}{2}, S_z = \frac{\rho_{bb}-\rho_{cc}}{2}. \label{eq:Sxyz}
\end{eqnarray}
The spin components also couple to the populations, $P_0=\rho_{aa}$, $P_1=\rho_{bb}+\rho_{cc}$ and $P_2=\rho_{dd}$. Instead of $\{\rho_{aa}, \rho_{bb}, \rho_{cc}, \Re[\rho_{bc}],\Im[\rho_{bc}], \rho_{dd}\}$, we now could have a new set of variables $\{P_0, P_1, S_z, S_x,S_y,P_2\}$. Then the master equation for density matrix elements is equivalent to the combination of rate equations for populations $P_0$, $P_1$, $P_2$ and Bloch-like equations for the spin components, see Appendix~\ref{sec:Bloch}.

\section{Fully symmetric two capacitively coupled quantum dots   
\label{sec:Qinfer}}

In this section we focus on the symmetric case of degenerate dot orbitals, $\delta \varepsilon = 0$, and identical tunnelling couplings, i.e., all $\delta_j = 0$ (see Fig.~\ref{fig:schematic_N2}). In this case, the commutator in Eq.~(\ref{eq:MBME}) is zero and Eqs.~(\ref{eq:Lindbladian_noInterfer}) and (\ref{eq:Lindbladian_yesInterfer}) simplify to Eqs.~(\ref{eq:Lindbladian_noInterfer_degen}) and (\ref{eq:Lindbladian_yesInterfer_degen}) in Appendix~\ref{sec:Append_Symmetric}.

\begin{figure}
\centering
  \includegraphics[width=.999\columnwidth]{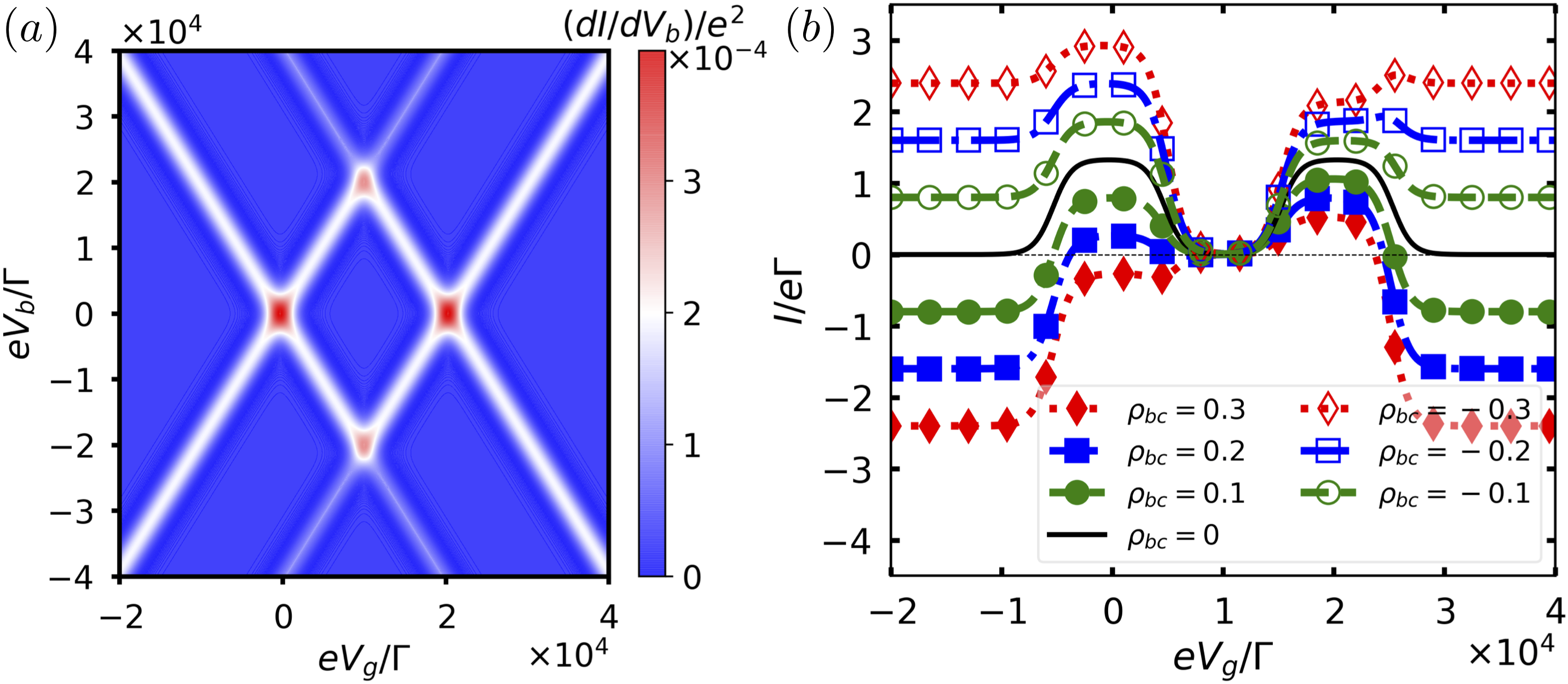} 
\caption{(Color online) Results for a completely symmetric double quantum dot.
(a) Stability diagram ($dI/dV_b$ as a function of $V_g$ and $V_b$) obtained by solving the Pauli rate equation for the diagonal elements of the density matrix. 
(b) Current at $eV_b = 10^4 \Gamma$ as a function of $V_g$ (along a horisontal cut in a stability diagram like in (a)), obtained by solving Eqs.~(\ref{eq:rho_bb_0}) and~(\ref{eq:rho_dd_0}) for different choices of $\rho_{bc}$.
The parameters used in both (a) and (b) are $U=2\times10^4\Gamma$, $\mu_{\rm L}=-\mu_{\rm R}=eV_b/2$, $T=862\Gamma$.}
\label{fig:analytics}
\end{figure}

We first present the result of solving the Pauli rate equations, which corresponds to including only the term $\mathcal{L}_{\square}\rho$ in Eq.~(\ref{eq:MBME}) and assuming the density matrix to be diagonal. The resulting stability diagram ($dI/dV_b$ as a function of $V_g$ and $V_b$) is shown in Fig.~\ref{fig:analytics}(a) and exhibits the typical Coulomb blockade behavior. At low $V_b$, the number of electrons on the double quantum dot is fixed and no current flows, except close to the two charge degeneracy points at $eV_g = 0$ and $eV_g = U$. Increasing $V_g$ starting from negative values, the charge on the double quantum dot is increased from zero to one and then to two electrons. At larger $V_b$, outside the Coulomb blockade region, a current flows because of single-electron tunneling.

We now turn to a solution for the full, possibly non-diagonal density matrix. For the fully symmetric case, setting the time derivates to zero in the equations in Appendices~\ref{sec:Append_Deriv} and~\ref{sec:Append_Symmetric} yields a singular matrix equation. This shows that there is no unique stationary state. For the time-dependent case, one would obtain a solution which, in principle, depends on the initial state regardless of how long one waits. 

The singular behavior can be understood by noticing that the equations for $\rho_{bb}$ and $\rho_{cc}$ are exactly the same, implying $\rho_{bb}=\rho_{cc}$. Because $\rho_{cb} = \rho_{bc}^{*}$ must hold, there are then only two independent equations for the three independent elements $\rho_{bb}, \rho_{bc}, \rho_{dd}$
\begin{eqnarray}
0 &=& 4\Gamma \sum_s [f_s(\varepsilon)\rho_{aa} 
- \bar{f}_s(\varepsilon) (\rho_{bb} + \rho_{bc}) ], \label{eq:rho_bb_0} \\
0 &=& 4\Gamma \sum_s [f_s(\varepsilon+U) (\rho_{bb}-\rho_{bc}) - \bar{f}_s(\varepsilon+U)\rho_{dd} ], \label{eq:rho_dd_0}
\end{eqnarray}
where $\rho_{aa}=1-2\rho_{bb}-\rho_{dd}$. One can thus fix one density matrix element, say $\rho_{bc}$, and then find a unique solution for the other elements. For example, the choice $\rho_{bc} = 0$ gives a solution without quantum interference which is the same as would be obtained by solving the Pauli rate equation for the diagonal density matrix elements. 

Figure~\ref{fig:analytics}(b) shows the current obtained by solving Eqs.~(\ref{eq:rho_bb_0}) and (\ref{eq:rho_dd_0}) with different choices for $\rho_{bc}$ and inserting the solution into Eq.~(\ref{eq:current}). The curve with $\rho_{bc}=0$ corresponds to the current along a horisontal cut at $eV_b = 10^4 \Gamma$ in Fig.~\ref{fig:analytics}(a). Compared with this solution, quantum interference ($\rho_{bc} \neq 0$) can either enhance or reduce the current. Note, however, that depending on the choice of $\rho_{bc}$ the results can become clearly unphysical, with negative occupation probabilities and even currents flowing against the applied bias voltage. 
We emphasize that the unphysical results originate from choosing $\rho_{bc}$ when the steady-state equations do not have a unique solution.
Importantly, the physical solutions we will obtain below in the presence of weak symmetry breaking effects never display such unphysical behavior. 
This is because the corresponding equations given below have well-determined steady-state solutions.

The singular behavior can alternatively be understood by reformulating the master equation in terms of the pseudo-spin components $S_x$, $S_y$, and $S_z$ [Eq.~(\ref{eq:Sxyz})], as well as the populations $P_0$, $P_1$, and $P_2$. The explicit equations are given in Appendix~\ref{sec:Bloch}. 
In the fully symmetric case, the equations for the spin components [Eqs.~(\ref{eq:Sx_asymm})--(\ref{eq:Sz_asymm})] become 
\begin{eqnarray}
\dot{S}_{x} &=& 
- 2 \Gamma \sum_s  [ \bar{f}_s(\varepsilon) + f_s(\varepsilon+U) ] S_x + \Gamma \sum_s \{ 2 f_s(\varepsilon) P_0 \notag\\
&& 
 - [ \bar{f}_s(\varepsilon) - f_s(\varepsilon+U) ] P_1 - 2 \bar{f}_s(\varepsilon+U) P_2 \} , \label{eq:Sx_symm} \\
\dot{S}_{y}&=& - 2 \Gamma \sum_s \{ [ \bar{f}_s(\varepsilon) + f_s(\varepsilon+U) ] S_y + B_x^s S_z\}, \label{eq:Sy_symm} \\
\dot{S}_z &=& 2\Gamma \sum_s \{ B_x^s S_y 
- [\bar{f}_s(\varepsilon) + f_s(\varepsilon+U)] S_z \}. \label{eq:Sz_symm}
\end{eqnarray}
Here,  $B_x^s=\frac{1}{\pi} [p_s(\varepsilon) + p_s(\varepsilon+U)]$ is an effective magnetic field in the $x$ direction [see Eq.~(\ref{eq:Lindbladian_principalpart})]. Equations~(\ref{eq:P0_asymm})--(\ref{eq:P2_asymm}) for the populations become 
\begin{eqnarray}
\dot{P}_0&=&2\Gamma \sum_s [-2f_s(\varepsilon) P_0 + \bar{f}_s(\varepsilon) P_1 + 2\bar{f}_s(\varepsilon) S_x ], \\
\dot{P}_{1}&=& 2\Gamma \sum_s \{2 f_s(\varepsilon) P_0
- [ \bar{f}_s(\varepsilon) +  f_s(\varepsilon+U)] P_1 \notag\\
&&+ 2 \bar{f}_s(\varepsilon+U) P_2 - 2 [\bar{f}_s(\varepsilon) -f_s(\varepsilon+U)] S_x  \}, \label{eq:P1_symm}\\
\dot{P}_{2}&=& 2 \Gamma \sum_s [ f_s(\varepsilon+U) P_1 - 2 \bar{f}_s(\varepsilon+U) P_2 \notag\\
&& - 2 f_s(\varepsilon+U) S_x ].  \label{eq:P2_symm}
\end{eqnarray}
Here, spin accumulation appears only in the $x$ direction because only the spin component $S_x$ depends on the populations. In fact, there are two sets of equations which do not couple to each other, one set for $\{S_x,P_1,P_0,P_2\}$ and one for $\{S_y,S_z\}$. In the stationary state, Eqs.~(\ref{eq:Sy_symm}) and (\ref{eq:Sz_symm}) imply that $S_y = S_z = 0$. Note that this also means that there is no dependence on the principle value integrals.

Similar to Eqs.~(\ref{eq:rho_bb_0}) and (\ref{eq:rho_dd_0}) for the density matrix, after using $P_0 + P_1 + P_2 = 1$, there are only two independent equations for the remaining three variables, for example:
\begin{eqnarray}
\dot{S}_x -\frac{1}{2}\dot{P}_1 &=& 
- 4 \Gamma \sum_s f_s(\varepsilon+U) S_x + \Gamma \sum_s \{ 2f_s(\varepsilon+U) P_1\notag\\
&& - 4 \bar{f}_s(\varepsilon+U) P_2 \}, 
\end{eqnarray}
and
\begin{eqnarray}
\dot{S}_x +\frac{1}{2}\dot{P}_1 &=&
- 4 \Gamma \sum_s \bar{f}_s(\varepsilon)  S_x + \Gamma \sum_s \{ 4 f_s(\varepsilon) P_0 \notag\\
&& - 2 \bar{f}_s(\varepsilon)  P_1  \} .
\end{eqnarray}
Therefore, we can solve the equations if we choose the value of one parameter. For example, fixing $S_x$ would correspond to fixing $\rho_{bc}$, and setting $S_x = 0$ leads to the same result as the Pauli rate equation.

Further insight into the singular nature of the master equation can be obtained by a change of basis for the states with one electron on the double quantum dot 
\begin{eqnarray}
|+\rangle &=& \frac{1}{\sqrt{2}} (|b\rangle + |c\rangle),\label{eq:plusstate}\\
|-\rangle &=& \frac{1}{\sqrt{2}} (|b\rangle - |c\rangle). \label{eq:minusstate}
\end{eqnarray}
These are analogous to the bonding and anti-bonding states of an electron in two degenerate orbitals, but they here have the same energy because there is no direct tunnel coupling between the orbitals. We now note that
\begin{eqnarray}
H_{\rm T} |+\rangle &=& \sum_{k,s={\rm L,R}} t_{k,s} c_{k,s}^{\dagger} \sqrt{2} |a\rangle, \label{eq:plusstatetunnel}\\
H_{\rm T} |-\rangle &=& -\sum_{k,s={\rm L,R}} t_{k,s}^* c_{k,s} \sqrt{2} |d\rangle. \label{eq:minusstatetunnel}
\end{eqnarray}
This shows that tunneling can only change the state of the double quantum dot either between $|a\rangle$ (no electrons on the double dot) and $|+\rangle$, or between $|d\rangle$ (two electrons on the double dot) and $|-\rangle$. Thus, in this basis it becomes clear that the master equation is separated into two different sectors which are not connected to each other through tunneling. Therefore, there can be no unique stationary state. 

\section{Symmetry-breaking effects \label{sec:pertur}}
We now consider small deviations away from perfect symmetry, i.e., $\delta \varepsilon$ and/or $\delta_j$ finite, but still much smaller than all other energy scales of the problem. Based on our conclusions for the symmetric system, a number of questions arise: Will arbitrarily small symmetry-breaking terms remove the singular behavior of the master equation and restore a well-defined unique stationary state? Will this stationary state be the same regardless of the details of the perturbations, as long as they are small? Will the possible stationary state(s) be one (or a subset) of the possible states for the symmetric system, or something different? 

In the following, we first consider two limiting cases, breaking either the orbital degeneracy or the tunnelling coupling symmetry. We then investigate the case where both symmetries are broken simultaneously. 

\subsection{Breaking orbital degeneracy}

We first consider symmetric tunnel couplings, $\delta_j = 0$, but a small breaking of the degeneracy of the quantum dot orbitals, $\delta \varepsilon \ll T, U, \Gamma$. We find that the stationary master equation (or, equivalently, the Block-like equation for the pseudo spin) becomes well defined for an arbitrarily small but finite $\delta \varepsilon$. Moreover, the stationary density matrix is fully diagonal and the current is equivalent to that given by the Pauli rate equation for the diagonal density matrix. The solution is equal to that in Fig.~\ref{fig:analytics}(a) and independent of $\delta \varepsilon$ as long as it remains by far the smallest energy scale. It is well known that the off-diagonal elements of the density matrix vanish for large orbital detuning ($\delta \varepsilon \gg \Gamma$), but the result that they are zero even for an arbitrarily small detuning only holds when the tunnel couplings are fully symmetric.

\begin{figure}[t]
\centering
  \includegraphics[width=1.0\columnwidth]{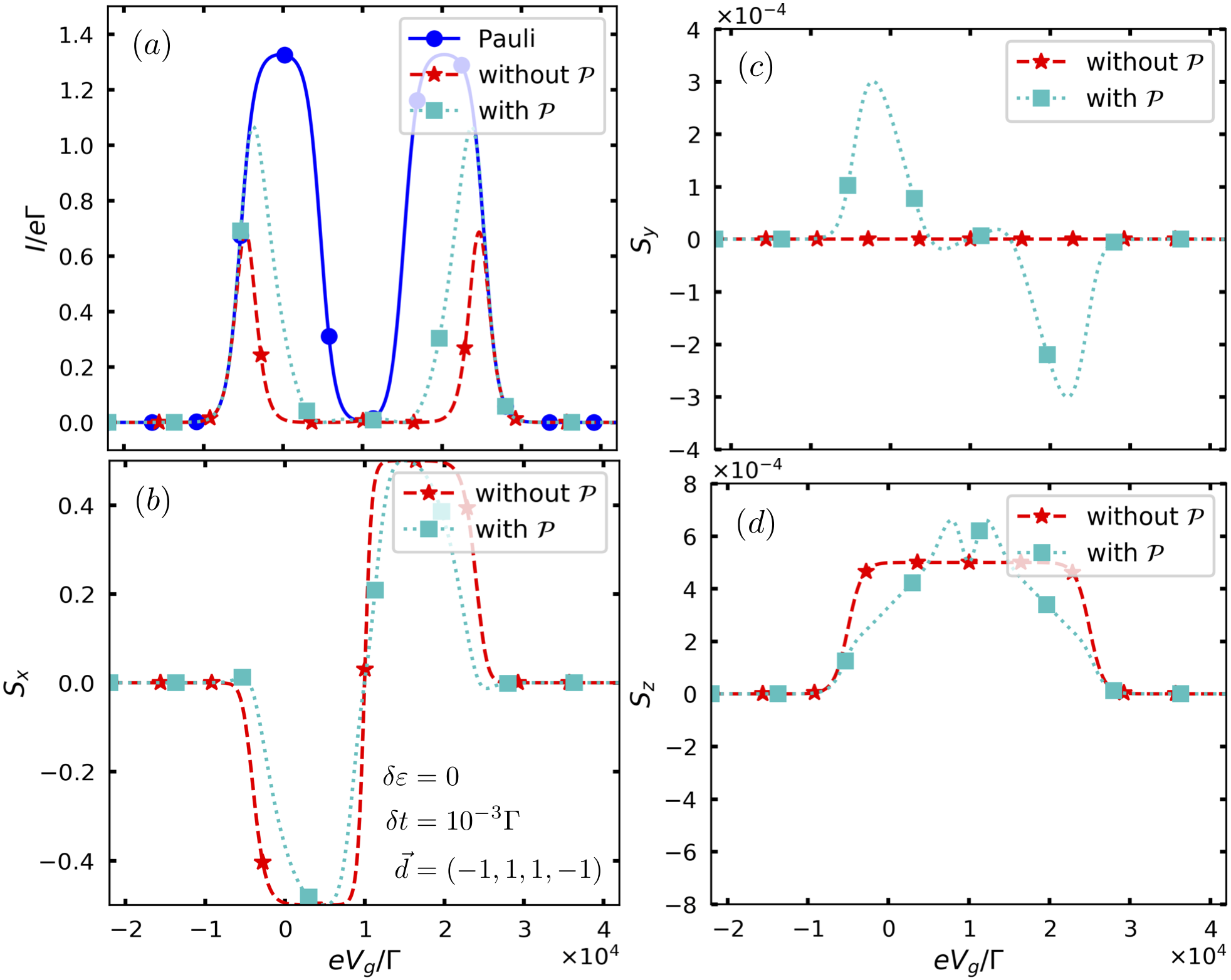} 
\caption{(Color online) Current and spin components as a function of $V_g$ obtained from different approximations to the master equation. The parameter are $\delta\varepsilon=0$, $\delta t=10^{-3}\Gamma$, $\vec{d}=(-1,1,1,-1)$, $U=2\times10^4\Gamma$, $eV_b=10^4\Gamma$, $T = 862\Gamma$.}
\label{fig:deltaepsilon0_I_Sx_Sy_Sz}
\end{figure}

\subsection{Breaking tunnel coupling symmetry}
We now turn to the case of complete orbital degeneracy, $\delta \varepsilon = 0$, but with some asymmetry in the tunnel couplings which we parametrize as (see Fig.~\ref{fig:schematic_N2}) $\delta_i = \delta t \times d_i$ with $d_i$ of order 1 and $\delta t \ll T, U, \Gamma$. We will consider different configurations of the tunnel coupling asymmetry, but note two special cases, $\vec{d} = (1,1,1,1)$ (equivalent to the symmetric case) and $\vec{d} = (d_0, d_1, d_0, d_1)$ (representing asymmetric left/right couplings, but equal couplings to both quantum dots). Except these two special cases, we find that an arbitrarily small tunnel coupling asymmetry leads to a well-defined master equation with a unique stationary state. The explicit equations for the pseudo-spin components and populations for the case $\delta \varepsilon = 0$, $\delta t \neq 0$ are presented in Appendix~\ref{sec:appendix_tunnelingasymmetry}.

Figure~\ref{fig:deltaepsilon0_I_Sx_Sy_Sz}(a) shows the current as a function of $V_g$ [similar to Fig.~\ref{fig:analytics}(b)] for $\vec{d} = (-1,1,1,-1)$ and $\delta t = 10^{-3} \Gamma$, comparing the solution for the Pauli rate equation and the master equation with and without including the principle value integrals. The stationary state is in this case not described by a diagonal density matrix. Quantum interference has a large impact on the current, which becomes suppressed to almost zero over a large range in $V_g$ where the Pauli rate equation predicts a large current. Interestingly, we find that the stationary density matrix and current are independent of $\delta t$ (as long as it remains much smaller than all other energy scales) and independent of $\vec{d}$ (except for the two special cases mentioned above). The results presented in Fig.~\ref{fig:deltaepsilon0_I_Sx_Sy_Sz} are therefore universal for the case $\delta \varepsilon = 0$ and $\delta t \ll T, U, \Gamma$.  

We can gain some understanding based on the states $|+\rangle$ and $|-\rangle$ defined in Eqs.~(\ref{eq:plusstate}) and~(\ref{eq:minusstate}). If we focus first on the $V_g$ range where current flow is associated with fluctuations between zero and one electron on the double dot [left peak in Fig.~\ref{fig:deltaepsilon0_I_Sx_Sy_Sz}(a)], Eqs.~(\ref{eq:plusstatetunnel}) and~(\ref{eq:minusstatetunnel}) show that for the completely symmetric case, transport can only involve fluctuations between states $|a\rangle$ and $|+\rangle$. Thus, $|+\rangle$ act as a bright state, while $|-\rangle$ becomes effectively decoupled from the leads and acts as a dark state. The occupation of the states $|+\rangle$ and $|-\rangle$ are given by
\begin{eqnarray}
\rho_{++}&=&\frac{1}{2}(\rho_{bb}+\rho_{cc}+\rho_{bc}+\rho_{cb}) = \frac{1}{2}P_1 + S_x, \label{eq:rhoplusplus} \\
\rho_{--}&=&\frac{1}{2}(\rho_{bb}+\rho_{cc}-\rho_{bc}-\rho_{cb}) = \frac{1}{2}P_1 - S_x. \label{eq:rhominusminus}
\end{eqnarray}
When now introducing a small symmetry-breaking term, this gives rise to a small coupling between the $|-\rangle$ state and the leads. In Fig.~\ref{fig:deltaepsilon0_I_Sx_Sy_Sz}(b) we see that this coupling results in $S_x$ attaining large negative values over a range in $V_g$ that precisely correspond to the suppression of the left current peak (compared with the Pauli rate equation). A large negative $S_x$ corresponds to a large occupation of the $|-\rangle$ state [Eq.~(\ref{eq:rhominusminus})] which acts as a dark state in this $V_g$ range. Thus, the current is decreased because the dark state, which is only very weakly coupled to the leads, becomes occupied with large probability and blocks transport through the bright state because the large Coulomb energy $U$ shifts the doubly occupied state $|d\rangle$ high up in energy.

In the $V_g$ range corresponding to the right current peak, where current flow is associated with fluctuations between one and two electrons on the double dot, Eqs.~(\ref{eq:plusstatetunnel}) and~(\ref{eq:minusstatetunnel}) show that instead the $|+\rangle$ state is the dark state, while $|-\rangle$ is the bright state. Here, the suppression of the current is instead associated with a significant positive $S_x$, meaning a large occupation of $|+\rangle$.

We can also understand the large impact of the principle value integrals (which are often neglected in master equations). Figure~\ref{fig:deltaepsilon0_I_Sx_Sy_Sz}(b) shows that the principle value integrals reduce $|S_x|$ over a range in $V_g$, leading to a larger current compared to the case where they are neglected. The reason for the decrease in $|S_x|$ is the pseudo magnetic field in Eqs.~(\ref{eq:Sx_delta_t})--(\ref{eq:Sz_delta_t}) which allows the spin to rotate away from the $x$ axis (or the electron to escape from the dark state). 

Figures~\ref{fig:deltaepsilon0_I_Sx_Sy_Sz}(c) and (d) show $S_y$ and $S_z$, which remain much smaller than $S_x$ for all $V_g$. 

\subsection{Breaking both orbital degeneracy and tunnel coupling symmetry}
We now turn to the general case with $\delta \varepsilon \neq 0$ and $\delta t \neq 0$, but still $\delta \varepsilon, \delta t \ll T, U, \Gamma$.  
\begin{figure*}
\centering
  \includegraphics[width=2\columnwidth]{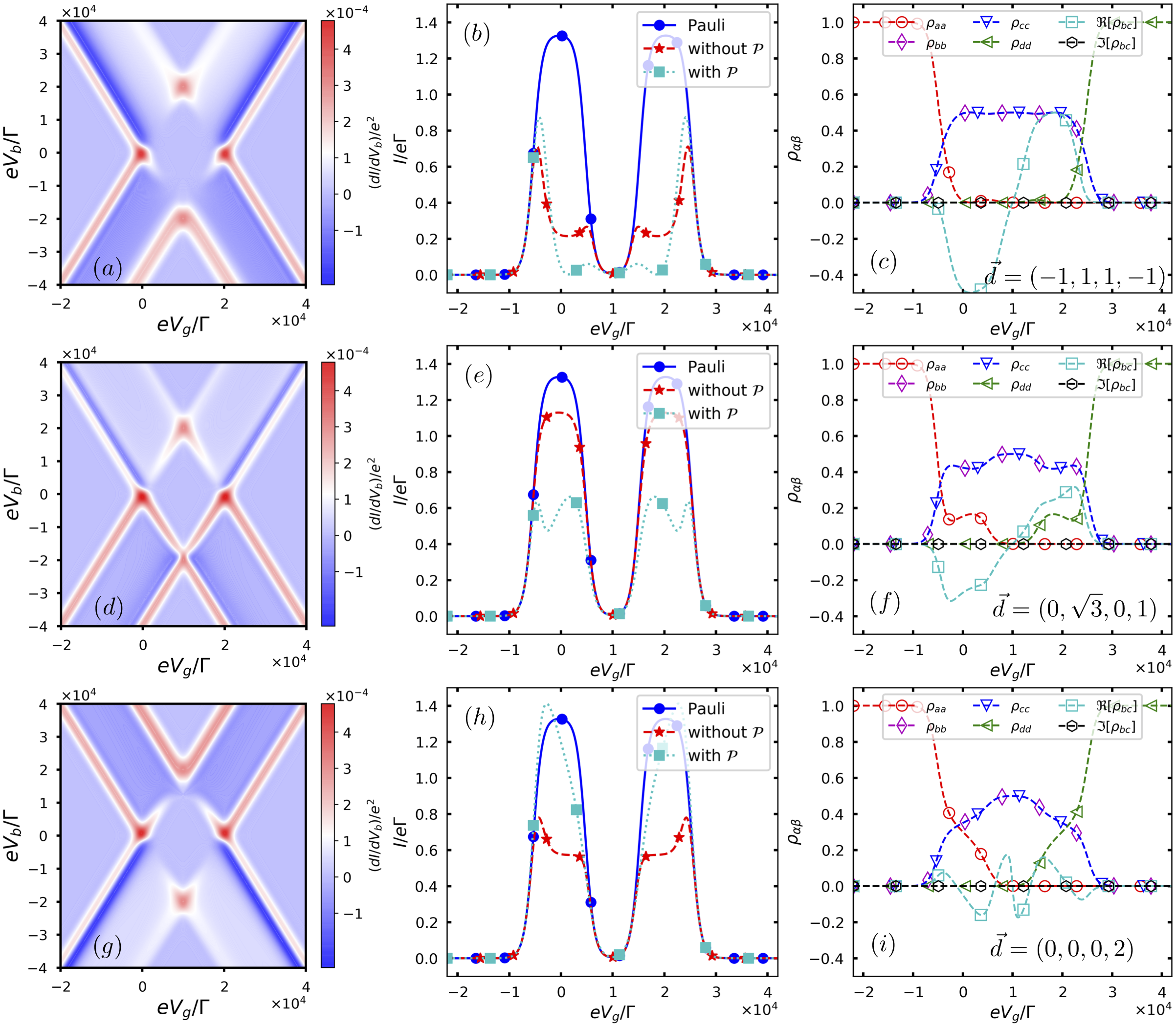} 
\caption{(Color online) Stability diagrams [(a), (d) and (g)], current as a function of $V_g$ with $eV_b = 10^4 \Gamma$ [(b), (e) and (h)] and density matrix elements as a function of $V_g$ with $eV_b = 10^4 \Gamma$ [(c), (f) and (i)]. $\vec{d}$ is varied while keeping the length of the vector fixed, $\vec{d} = (-1,1,1,-1)$ in (a)--(c), $\vec{d} = (0,\sqrt{3},0,1)$ in (d)--(f) and $\vec{d} = (0,0,0,2)$ in (g)--(i). Here $\delta t=\delta\varepsilon=10^{-3}\Gamma$ and all other parameters are the same as in Fig.~\ref{fig:deltaepsilon0_I_Sx_Sy_Sz}.}
\label{fig:elec_deltaVg_asymmetry}
\end{figure*}
Figure~\ref{fig:elec_deltaVg_asymmetry} shows the stability diagrams, current and density matrix elements as a function of $V_g$ for three different choices of $\vec{d}$ (meaning different configurations of the asymmetry in tunnel couplings) for $\delta t = \delta \varepsilon = 10^{-3} \Gamma$. The current clearly deviates from that given by the Pauli rate equation and, unlike the case with $\delta \varepsilon = 0$, depends on the specific choice of $\vec{d}$ (the result of the Pauli rate equation is, in contrast, independent of $\vec{d}$ as long as $\delta t \ll T, U, \Gamma$). We also note that including the principle value integrals has a large impact on the current, which can be either enhanced or suppressed compared with the (commonly used) master equation where these terms are neglected. 

Interestingly, the stability diagrams in Figs.~\ref{fig:elec_deltaVg_asymmetry}(a), (d) and (g) all show regions of negative differential resistance, as well as strong rectifying behavior (different magnitude of the current for positive and negative $V_b$). Such effects are expected in quantum dot systems when the tunnel couplings to different orbitals and/or leads differ substantially, but here they are induced by quantum interference even in the case where the system is very close to symmetric.

%\begin{figure}%[t]
%\centering
%  \includegraphics[width=.999\columnwidth]{analytics_pertur.png} 
%\caption{(colour online) The electrical current through double quantum dots  
%under perturbations (e.g., nonzero $\delta\varepsilon$ and $\delta t$) for (a) $\vec{d}=\{-1,1,1,-1\}$ and (b) $\vec{d}=\{1,1,1,1\}$ at $eV_b=10^3\Gamma$ as a function of the gate voltage $eV_g$. The differential conductance (c) and (d) for $\delta\varepsilon=\delta t=10^{-3}\Gamma$ corresponding to the solid curves in (a) and (b) respectively. Other parameters are same as those in Fig. \ref{fig:analytics}.}
%\label{fig:analytics_pertur}
%\end{figure}

\begin{figure}
\centering
  \includegraphics[width=.999\columnwidth]{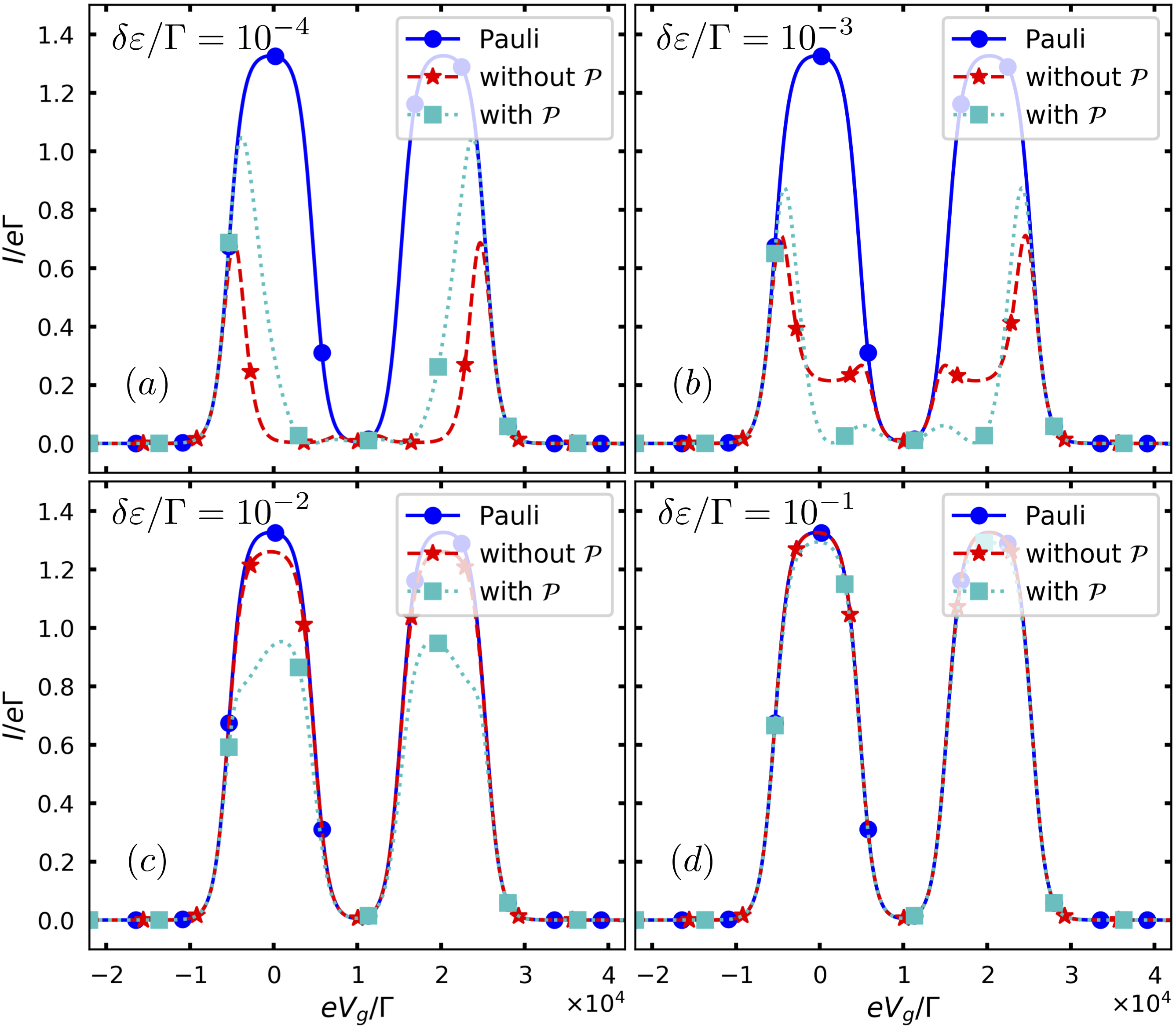} 
\caption{(Color online) Current as a function of $V_g$ with $eV_b = 10^4 \Gamma$ for different values of $\delta \varepsilon$ at fixed $\delta t=10^{-3}\Gamma$ and $\vec{d}=(-1,1,1,-1)$. All other parameters are the same as in Fig.~\ref{fig:deltaepsilon0_I_Sx_Sy_Sz}.}
\label{fig:N2_electric_Vg_delta_asymm}
\end{figure}

\begin{figure}%[t]
\centering
  \includegraphics[width=.9\columnwidth]{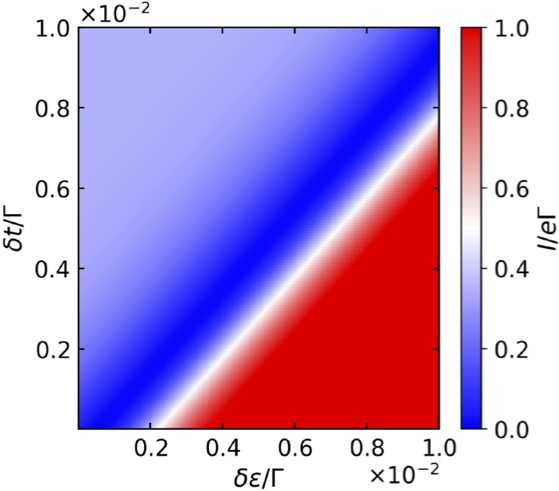} 
\caption{(Color online) Current as a function of $\delta\varepsilon$ and $\delta t$ at fixed $eV_g=0$ with $eV_b = 10^4 \Gamma$. All other parameters are the same as in Fig.~\ref{fig:deltaepsilon0_I_Sx_Sy_Sz}.}
\label{fig:competition}
\end{figure}

Figure~\ref{fig:N2_electric_Vg_delta_asymm} shows the current as a function of $V_g$ for fixed $\delta t$ and $\vec{d}$, but with increasing $\delta \varepsilon$. For $\delta \varepsilon \ll \delta t$ the current is suppressed over a large range in $V_g$, similar to Fig.~\ref{fig:analytics}. The current starts to recover when $\delta \varepsilon \sim \delta t$ and approaches the value given by the Pauli rate equation for $\delta \varepsilon \gg \delta t$ (even though $\delta \varepsilon \ll \Gamma$ still holds). As long as $\delta \varepsilon, \delta t \ll T, U, \Gamma$ holds, the current depends only on the ratio $\delta \varepsilon / \delta t$ (and on $\vec{d}$ if $\delta \varepsilon \sim \delta t$). The dependence on $\delta \varepsilon / \delta t$ is shown in Fig.~\ref{fig:competition} where we plot the current as a function of $\delta \varepsilon$ and $\delta t$ for fixed $V_g = 0$. The current switches between its two limiting values along a diagonal line, which shows that very small controlled changes in $\delta \varepsilon$ or $\delta t$ can have a large impact on the current. 

\section{Summary and conclusions \label{sec:discuss}}
In this work, we have shown that for a double quantum dot with degenerate orbitals and equal tunnel couplings to both dots and both leads, the master equation for the stationary density matrix becomes singular. This singular behavior implies that a whole family of stationary states is allowed by the equations, in principle, one of them being the diagonal density matrix as obtained from the Pauli rate equation. When including perturbations away from the perfectly symmetric system, the stationary state becomes unique. When considering only breaking of orbital degeneracy (but symmetric tunnel couplings), this stationary state is a diagonal density matrix, even if the orbital detuning is much smaller than all other energy scales. In the opposite case of orbital degeneracy but small perturbations to the tunnel couplings, the density matrix is non-diagonal and the current is significantly suppressed by quantum interference. In the case of breaking both orbital degeneracy and tunnel coupling symmetry, the current is a very sensitive function of the ratio of the symmetry breaking terms. 

We note that our study assumes that both quantum dots are coupled to the same lead states. This is a good assumption roughly when the points where the two dots couple to a given lead are separeted by less than the Fermi wave length in the leads. This condition can be realized with semiconductor leads (large Fermi wavelength) or by using a molecular double quantum dot (small distance between dots). The effects of increasing the distance between the quantum dots were investigated in~\onlinecite{Kubo06prb, Tokura07njp}, where it was shown to destroy quantum interference features. 

We have relied on a master equation which is based on leading order perturbation theory in the tunnel couplings, which is expected to be a very good approximation in all cases we consider because we stay in the regime $\Gamma \ll T$. Nonetheless, we have used the methods described in Refs.~\onlinecite{qmeq17, Leijnse2008} to verify that processes which are next-to-leading order in the tunneling coupling do not affect our conclusions. It would be interesting to, in a future study, investigate the regime $\Gamma \lesssim T$ where cotunneling and other higher order tunnel processes become important. 
Note that our Hamiltonian of two capacitively coupled quantum dots in terms of spinless fermionic operators resembles a Hubbard site for spinful electrons. It is therefore possible to reformulate the model considered in the present work in terms of a single dot with spin up and spin down electrons coupled to partially spin-polarized leads, as implied by a pseudo spin defined in Eq.~(\ref{eq:Sxyz}). For strong tunnel couplings, one would expect the appearance of Kondo effect associated with the pseudo spin. In the present work we focus instead on the regime of weak tunnel coupling.
It might also be interesting to consider shot noise or full counting statistics that could provide more information about electron transport in our system.~\cite{Nazarov03} 

Decoherence effects due to charge noise or phonons, not included in this study, would lead to a decay of electrons trapped in the dark state. However, the qualitative results are not affected by weak decoherence. We note that a recent demonstration of Landau-Zener-St\"{u}ckelberg-Majorana interferometry in a silicon-based single-electron double quantum dot suggests that decoherence can be weak enough for quantum interference to survive~\cite{Chatterjee18prb}. 

From an application perspective, the double quantum dot system we investigate could be used as an extremely sensitive electric switch, similar to a field effect transistor. If an additional electrostatic gate is used to control the tunnel coupling asymmetry ($\delta t$) or the orbital detuning ($\delta \varepsilon$), one can switch the current between almost zero and a large finite value by a tiny shift in either of these parameters which makes the system cross the diagonal in Fig.~\ref{fig:competition}. Note that the change happens when changing a parameter by an amount that is much smaller than all other energy scales. In particular, unlike most other switches, this switch will not be limited by temperature.

\acknowledgements

We acknowledge helpful discussions with Martin Josefsson and Andreas Wacker, and financial support from the Carl Trygger Foundation and from NanoLund. 

\appendix
\section{Explicit form of the Master equation\label{sec:Append_Deriv}}
Starting from Eq.~(\ref{eq:MBME}), the explicit forms of Eqs.~(\ref{eq:Lindbladian_noInterfer})--(\ref{eq:Lindbladian_principalpart}) for the different density matrix elements are   
\begin{eqnarray}
\dot{\rho}_{aa}&=&\sum_s\{
 -2 [\Gamma_{s1} f_s(\varepsilon_1) + \Gamma_{s2} f_s(\varepsilon_2)] \rho_{aa} + 2 \Gamma_{s1} \bar{f}_s(\varepsilon_1) \rho_{bb}   \notag\\
&&
+ 2 \Gamma_{s2} \bar{f}_s(\varepsilon_2) \rho_{cc} \} + \sum_s\sqrt{\Gamma_{s1}\Gamma_{s2}} \{ [ \bar{f}_s (\varepsilon_1) + \bar{f}_s(\varepsilon_2) ] \notag\\
&& \times ( \rho_{cb}  +  \rho_{bc}  )  + \frac{i}{\pi}[p_s(\varepsilon_1)-p_s(\varepsilon_2)] (\rho_{bc} - \rho_{cb} ) \}, 
\label{eq:rho_aa} 
\end{eqnarray}
\begin{eqnarray}
\dot{\rho}_{bb}&=&
\sum_s\{
2 \Gamma_{s1} f_s(\varepsilon_1) \rho_{aa}
-2 [\Gamma_{s1}  \bar{f}_s(\varepsilon_1) + \Gamma_{s2}f_s(\varepsilon_2+U)] \notag\\
&& \times \rho_{bb} 
+2 \Gamma_{s2} \bar{f}_s(\varepsilon_2+U) \rho_{dd} \} - \sum_s\sqrt{\Gamma_{s1}\Gamma_{s2}} \{ [\bar{f}_s(\varepsilon_2) \notag\\
&& - f_s(\varepsilon_1+U) ] ( \rho_{cb} +  \rho_{bc} ) - \frac{i}{\pi} [p_s(\varepsilon_2)+p_s(\varepsilon_1+U)] \notag\\
&& \times ( \rho_{bc} - \rho_{cb})] \} , 
\label{eq:rho_bb} 
\end{eqnarray}
\begin{eqnarray}
\dot{\rho}_{cc}&=&
\sum_s \{
2 \Gamma_{s2} f_s(\varepsilon_2) \rho_{aa}
-2 [ \Gamma_{s2} \bar{f}_s(\varepsilon_2) + \Gamma_{s1} f_s(\varepsilon_1+U)] \notag\\
&& \times  \rho_{cc}
+2 \Gamma_{s1} \bar{f}_s (\varepsilon_1+U) \rho_{dd} \} - \sum_s\sqrt{\Gamma_{s1}\Gamma_{s2}} \{ [ \bar{f}_s(\varepsilon_1) \notag\\
&&- f_s(\varepsilon_2+U) ] ( \rho_{bc} +  \rho_{cb} ) +\frac{i}{\pi} [p_s(\varepsilon_2+U) + p_s(\varepsilon_1)] \notag\\
&& \times (\rho_{bc} - \rho_{cb}) \}, 
\label{eq:rho_cc} 
\end{eqnarray}
\begin{eqnarray}
\dot{\rho}_{dd}&=&
\sum_s  \{
2 \Gamma_{s2} f_s(\varepsilon_2+U) \rho_{bb} 
+ 2 \Gamma_{s1}  f_s(\varepsilon_1+U) \rho_{cc} \notag\\
&& - 2 [ \Gamma_{s1}  \bar{f}_s(\varepsilon_1+U) + \Gamma_{s2} \bar{f}_s(\varepsilon_2+U)] \rho_{dd} \}  \notag\\
&&-\sum_s \sqrt{\Gamma_{s1}\Gamma_{s2}} \{ 
 [ f_s(\varepsilon_1+U) + f_s(\varepsilon_2+U) ] 
( \rho_{bc}  \notag\\
&& %- \sum_s\sqrt{\Gamma_{s1}\Gamma_{s2}} 
+ \rho_{cb} ) + \frac{i}{\pi} [ p_s(\varepsilon_1+U) - p_s(\varepsilon_2+U) ] (\rho_{bc} - \rho_{cb}) \}
, \label{eq:rho_dd} \notag\\
\end{eqnarray}
\begin{eqnarray}
\dot{\rho}_{bc}&=& i(\varepsilon_2-\varepsilon_1)\rho_{bc} +
\sum_s 
\sqrt{\Gamma_{s1}\Gamma_{s2}} \{ [ f_s(\varepsilon_1) + f_s(\varepsilon_2) ] \rho_{aa}  \notag\\
&& - [ \bar{f}_s(\varepsilon_1) - f_s(\varepsilon_2+U)] \rho_{bb} 
 - [ \bar{f}_s(\varepsilon_2) - f_s(\varepsilon_1+U) ]  \notag\\
&& \times \rho_{cc} - [ \bar{f}_s(\varepsilon_1+U) + \bar{f}_s(\varepsilon_2+U)] \rho_{dd} \}   - \sum_s \{ \Gamma_{s1} \notag\\
&& \times [ \bar{f}_s(\varepsilon_1) + f_s(\varepsilon_1+U) ]   
+\Gamma_{s2} [ \bar{f}_s(\varepsilon_2) + f_s(\varepsilon_2+U) ] \} \notag\\
&& \times  \rho_{bc} 
+  \sum_s \frac{i}{\pi} \{ \Gamma_{s2} [ p_s(\varepsilon_2) + p_s(\varepsilon_2+U)] - \Gamma_{s1}  \notag\\
&& \times [ p_s(\varepsilon_1) + p_s(\varepsilon_1+U)] \}  \rho_{bc}  + \sum_s \sqrt{\Gamma_{s1}\Gamma_{s2}}  \frac{i}{\pi}  \notag\\
&&\times  \{ [ p_s(\varepsilon_2) - p_s(\varepsilon_1) ]  \rho_{aa}  
+ [ p_s(\varepsilon_1) + p_s(\varepsilon_2+U)]  \notag\\
&&\times \rho_{bb}  -  [p_s(\varepsilon_1+U) + p_s(\varepsilon_2)] \rho_{cc} 
+ [p_s(\varepsilon_1+U) \notag\\
&&- p_s(\varepsilon_2+U)] \rho_{dd} \}.
\label{eq:rho_bc} 
\end{eqnarray}
From these expressions one can verify that $\sum_{\alpha=a,b,c,d} \dot{\rho}_{\alpha\alpha}=0$, which is due to probability normalization, $\sum_{\alpha=a,b,c,d} \rho_{\alpha\alpha}=1$.

\section{Master equation for the fully symmetric system\label{sec:Append_Symmetric}}
In the fully symmetric case with $\delta \varepsilon = 0$ and all $\delta_j = 0$, Eqs.~(\ref{eq:Lindbladian_noInterfer})--(\ref{eq:Lindbladian_principalpart})  become
\begin{eqnarray}
\mathcal{L}_{\square}\rho&=&\sum_{s={\rm L,R}} \Gamma_{s} \{ f_s(\varepsilon) (
\mathcal{D}[|b\rangle\langle a|] \rho + \mathcal{D}[|c\rangle\langle a|] \rho )  \notag\\
&&
+ \bar{f}_s(\varepsilon) ( \mathcal{D}[|a\rangle\langle b|] \rho +\mathcal{D}[|a\rangle\langle c|] \rho ) \notag\\
&&+ f_s(\varepsilon+U)
( \mathcal{D}[|d\rangle\langle c|] \rho +\mathcal{D}[|d\rangle\langle b|] \rho )  \notag\\
&&+ \bar{f}_s(\varepsilon+U) ( \mathcal{D}[|c\rangle\langle d|] \rho
+\mathcal{D}[|b\rangle\langle d|] \rho ) \} , \label{eq:Lindbladian_noInterfer_degen} 
\end{eqnarray} 

\begin{eqnarray}
\mathcal{L}_{\boxtimes}\rho&=&\sum_{s={\rm L,R}} \Gamma_{s} \{ f_s(\varepsilon)
\mathds{D}[|b\rangle\langle a|,|c\rangle\langle a|] \rho  \notag\\
&&+ \bar{f}_s(\varepsilon) \mathds{D}[|a\rangle\langle b|, |a\rangle\langle c|] \rho \notag\\
&&+ f_s(\varepsilon+U) \mathds{D}[|d\rangle\langle c|,|d\rangle\langle b|] \rho  \notag\\
&&+ \bar{f}_s(\varepsilon+U) \mathds{D}[|c\rangle\langle d|, |b\rangle\langle d|] \rho \},
\label{eq:Lindbladian_yesInterfer_degen}
\end{eqnarray} 
\begin{eqnarray}
\mathcal{P}\rho
&=& \frac{i}{\pi}
\sum_{s={\rm L,R}} \Gamma_{s} 
\{ p_s(\varepsilon) ([|a\rangle\langle a|-|b\rangle\langle b|,\rho] \notag\\
&&+ [|a\rangle\langle a|-|c\rangle\langle c|,\rho] 
+ [|a\rangle\langle c|,[|b\rangle\langle a|,\rho]] \notag\\
&&-[|c\rangle\langle a|,[|a\rangle\langle b|,\rho]] 
+ [|a\rangle\langle b|,[|c\rangle\langle a|,\rho]] \notag\\
&&-[|b\rangle\langle a|,[|a\rangle\langle c|,\rho]]) 
-p_s(\varepsilon+U) \notag\\
&&\times ([|c\rangle\langle c|-|d\rangle\langle d|,\rho] 
+ [|b\rangle\langle b|-|d\rangle\langle d|,\rho] \notag\\
&& + [|b\rangle\langle d|,[|d\rangle\langle c|,\rho]]-[|d\rangle\langle b|,[|c\rangle\langle d|,\rho]] \notag\\
&& + [|c\rangle\langle d|,[|d\rangle\langle b|,\rho]]-[|d\rangle\langle c|,[|b\rangle\langle d|,\rho]])
\}.
\end{eqnarray}
Here $\mathds{D}[A,B]\rho$ is the sum of two different tunneling pathways, which are equivalent due to the exact orbital degeneracy
\begin{eqnarray}
\mathds{D}[A,B]\rho&=& \mathcal{D}[A,B]\rho + \mathcal{D}[B,A]\rho \notag\\
&=& 2A\rho B^{\dagger}+2B\rho A^{\dagger} 
-\rho B^{\dagger}A -B^{\dagger}A\rho \notag\\
&& -\rho A^{\dagger}B -A^{\dagger}B\rho \label{eq:}.
\end{eqnarray} 
 
\section{Bloch-like equations for the pseudo spin\label{sec:Bloch}}

The Bloch-like equations for $S_x,S_y,S_z$ are 
\begin{eqnarray}
\dot{S}_{x} &=& 
-\sum_s \{ \Gamma_{s1} [ \bar{f}_s(\varepsilon_1) + f_s(\varepsilon_1+U) ] +\Gamma_{s2} [ \bar{f}_s(\varepsilon_2) \notag\\
&& + f_s(\varepsilon_2+U) ] \}  S_x + [(\varepsilon_2-\varepsilon_1) + \sum_s ( \Gamma_{s2} B_{z2}^s  \notag\\
&& %\frac{1}{\pi} [ p(\varepsilon_2) + p(\varepsilon_2+U)]  \notag\\
- \Gamma_{s1} B_{z1}^s %\frac{1}{\pi} [ p(\varepsilon_1) + p(\varepsilon_1+U)] 
) ]S_y
 - \sum_s \sqrt{\Gamma_{s1}\Gamma_{s2}} [ \bar{f}_s(\varepsilon_1) - f_s(\varepsilon_2+U) \notag\\
&& + f_s(\varepsilon_1+U) - \bar{f}_s(\varepsilon_2) ] S_z + \sum_s \sqrt{\Gamma_{s1}\Gamma_{s2}} \{ [ f_s(\varepsilon_1) \notag\\
&&  + f_s(\varepsilon_2) ] P_0 
- [ \bar{f}_s(\varepsilon_1+U) + \bar{f}_s(\varepsilon_2+U)] P_2 - [ \bar{f}_s(\varepsilon_1) \notag\\
&& - f_s(\varepsilon_2+U) - f_s(\varepsilon_1+U) + \bar{f}_s(\varepsilon_2) ]  \frac{P_1}{2} \} , 
\label{eq:Sx_asymm}
\end{eqnarray}
\begin{eqnarray}
\dot{S}_{y}&=& [ (\varepsilon_1 - \varepsilon_2) %S_x 
+ \sum_s ( \Gamma_{s1} B_{z1}^s %\frac{1}{\pi} [ p(\varepsilon_1) + p(\varepsilon_1+U)] \notag\\
%&&
- \Gamma_{s2} B_{z2}^s %\frac{1}{\pi} [ p(\varepsilon_2) + p(\varepsilon_2+U)] 
) ] S_x - \sum_s \{ \Gamma_{s1}  \notag\\
&& \times [ \bar{f}_s(\varepsilon_1) + f_s(\varepsilon_1+U) ] +\Gamma_{s2} [ \bar{f}_s(\varepsilon_2) + f_s(\varepsilon_2+U) ] \} \notag\\
&& \times S_y  - \sum_s\sqrt{\Gamma_{s1}\Gamma_{s2}} 
(B_{x1}^s+B_{x2}^s)%\frac{1}{\pi} [p(\varepsilon_1) + p(\varepsilon_2) + p(\varepsilon_1+U) + p(\varepsilon_2+U) ] 
S_z - \sum_s\sqrt{\Gamma_{s1}\Gamma_{s2}} \notag\\
&& \times \frac{1}{\pi} \{
[ p_s(\varepsilon_2)-p_s(\varepsilon_1) ] P_0 + [ p_s(\varepsilon_1+U) - p_s(\varepsilon_2+U)] P_2  \notag\\
&& + [ p_s(\varepsilon_1) - p_s(\varepsilon_1+U) + p_s(\varepsilon_2+U) - p_s(\varepsilon_2) ] \frac{P_1}{2} \},
\label{eq:Sy_asymm}
\end{eqnarray}
\begin{eqnarray}
\dot{S}_z &=&
\sum_s \sqrt{\Gamma_{s1}\Gamma_{s2}} \{
%[ \bar{f}_s(\varepsilon_1) - f_s(\varepsilon_2+U) \notag\\
%&& + f_s(\varepsilon_1+U) - \bar{f}_s(\varepsilon_2) ] 
(B_{y1}^s-B_{y2}^s) S_x 
 + %\sum_s \sqrt{\Gamma_{s1}\Gamma_{s2}} 
(B_{x1}^s+B_{x2}^s)%\frac{1}{\pi} [p(\varepsilon_1) + p(\varepsilon_2) + p(\varepsilon_1+U) + p(\varepsilon_2+U)] 
S_y \} \notag\\
&& - \sum_s [ \Gamma_{s2} \bar{f}_s(\varepsilon_2) + \Gamma_{s1} f_s(\varepsilon_1+U) +\Gamma_{s1}  \bar{f}_s(\varepsilon_1) \notag\\
&& + \Gamma_{s2}f_s(\varepsilon_2+U)] S_z 
 +\sum_s \{ [\Gamma_{s1} f_s(\varepsilon_1) - \Gamma_{s2} f_s(\varepsilon_2)]  \notag\\
&&\times P_0 + [\Gamma_{s2} \bar{f}_s(\varepsilon_2+U) - \Gamma_{s1} \bar{f}_s (\varepsilon_1+U)] P_2 \}  
 \notag\\
&& + \sum_s \{ \Gamma_{s2}  [\bar{f}_s(\varepsilon_2) - f_s(\varepsilon_2+U)] - \Gamma_{s1} [ \bar{f}_s(\varepsilon_1) \notag\\
&&  - f_s(\varepsilon_1+U) ] \} \frac{P_1}{2}.
\label{eq:Sz_asymm}
\end{eqnarray}
Here we defined the pseudo magnetic fields
\begin{eqnarray}
B_{xi}^{s} &=& B_{zi}^{s} = \frac{1}{\pi}[p_s(\varepsilon_i) + p_s(\varepsilon_i+U)], \\
%B_{x2}^{s} &=& B_{z2}^{s} = \frac{1}{\pi}[p_s(\varepsilon_2) + p_s(\varepsilon_2+U)], \\
B_{yi}^{s} &=& \bar{f}_s(\varepsilon_i) + f_s(\varepsilon_i+U), 
%B_{y2}^{s} &=& \bar{f}_s(\varepsilon_2) + f_s(\varepsilon_2+U),
\end{eqnarray}
where $ p_{s}(\omega) = \Re[\Psi (\frac{1}{2} + \frac{i}{2\pi} \frac{\omega-\mu_s}{T_s})]$ (see main paper). Note that the pseudo magnetic field in the $y$ direction vanishes in Eqs.~(\ref{eq:Sx_asymm})--(\ref{eq:Sy_asymm}) when $\delta\varepsilon=0$, while the pseudo magnetic field in the $z$ direction vanishes only when $\delta t = \delta \varepsilon = 0$. 

The rate equations for the populations read
\begin{eqnarray}
\dot{P}_0 &=&
\sum_s \{-2 [\Gamma_{s1} f_s(\varepsilon_1) + \Gamma_{s2} f_s(\varepsilon_2)] P_0 + [ \Gamma_{s1} \bar{f}_s(\varepsilon_1) \notag\\
&& + \Gamma_{s2} \bar{f}_s(\varepsilon_2) ] P_1 \} + \sum_s 2 \sqrt{\Gamma_{s1}\Gamma_{s2}} \{ [ \bar{f}_s (\varepsilon_1) \notag\\
&&  + \bar{f}_s(\varepsilon_2) ]  S_x 
+\frac{1}{\pi}[p_s(\varepsilon_1)-p_s(\varepsilon_2)] S_y \} \notag\\
&&+ \sum_s 2 [ \Gamma_{s1} \bar{f}_s(\varepsilon_1) - \Gamma_{s2} \bar{f}_s(\varepsilon_2) ] S_z , 
\label{eq:P0_asymm}
\end{eqnarray}
\begin{eqnarray}
\dot{P}_1 &=& 
\sum_s \{ 2 [\Gamma_{s1} f_s(\varepsilon_1) +\Gamma_{s2} f_s(\varepsilon_2)] P_0 +2 [\Gamma_{s2} \bar{f}_s(\varepsilon_2+U)  \notag\\
&&+ \Gamma_{s1} \bar{f}_s (\varepsilon_1+U)] P_2  
- [\Gamma_{s1}  \bar{f}_s(\varepsilon_1) + \Gamma_{s2}f_s(\varepsilon_2+U) \notag\\
&&+\Gamma_{s2} \bar{f}_s(\varepsilon_2) + \Gamma_{s1} f_s(\varepsilon_1+U)] P_1 \} - \sum_s 2 \sqrt{\Gamma_{s1}\Gamma_{s2}}  \notag\\
&&\times \{ [ \bar{f}_s(\varepsilon_1) - f_s(\varepsilon_1+U) 
+ \bar{f}_s(\varepsilon_2) - f_s(\varepsilon_2+U)] S_x  \notag\\
&& %+ \sum_s 2 \sqrt{\Gamma_{s1}\Gamma_{s2}} 
-\frac{1}{\pi} [p_s(\varepsilon_2) -p_s(\varepsilon_2+U) 
- p_s(\varepsilon_1) + p_s(\varepsilon_1+U)] S_y \}\notag\\
&& - \sum_s 2 [\Gamma_{s1}  \bar{f}_s(\varepsilon_1) + \Gamma_{s2}f_s(\varepsilon_2+U) -\Gamma_{s2} \bar{f}_s(\varepsilon_2) \notag\\
&& - \Gamma_{s1} f_s(\varepsilon_1+U)] S_z , 
\label{eq:P1_asymm}
\end{eqnarray}
\begin{eqnarray}
\dot{P}_{2} &=&
\sum_s \{ [\Gamma_{s1}  f_s(\varepsilon_1+U) + \Gamma_{s2} f_s(\varepsilon_2+U)] P_1 - 2 [ \Gamma_{s1}  \notag\\
&& \times \bar{f}_s(\varepsilon_1+U) + \Gamma_{s2} \bar{f}_s(\varepsilon_2+U)] P_2 \} - \sum_s 2 \sqrt{\Gamma_{s1}\Gamma_{s2}} \notag\\
&& \times \{ [ f_s(\varepsilon_1+U) + f_s(\varepsilon_2+U) ] S_x -\frac{1}{\pi} [ p_s(\varepsilon_2+U)  \notag\\
&& - p_s(\varepsilon_1+U) ] S_y\} + \sum_s 2 [ \Gamma_{s2} f_s(\varepsilon_2+U) \notag\\
&& - \Gamma_{s1}  f_s(\varepsilon_1+U)] S_z .
\label{eq:P2_asymm}
\end{eqnarray}

\subsection{Orbital degeneracy \label{sec:appendix_tunnelingasymmetry}}

When considering $\delta t\neq0$ but $\delta \varepsilon = 0$, the equations for the spin components reduce to
\begin{eqnarray}
\dot{S}_{x} &=&
- \sum_s (\Gamma_{s1} + \Gamma_{s2} ) [ \bar{f}_s(\varepsilon) + f_s(\varepsilon+U) ] S_x  + \sum_s (\Gamma_{s2}  \notag\\
&& - \Gamma_{s1}) 
\frac{1}{\pi}[ p_s(\varepsilon) + p_s(\varepsilon+U)] S_y + \sum_s \sqrt{\Gamma_{s1}\Gamma_{s2}}  \{ 2 f_s(\varepsilon) \notag\\
&& \times P_0 - [ \bar{f}_s(\varepsilon) - f_s(\varepsilon+U) ] P_1 - 2 \bar{f}_s(\varepsilon+U) P_2 \} ,  \label{eq:Sx_delta_t}
\end{eqnarray}
\begin{eqnarray}
\dot{S}_{y}&=& \sum_s (\Gamma_{s1} - \Gamma_{s2}) 
\frac{1}{\pi} [ p_s(\varepsilon) + p_s(\varepsilon+U)] S_x - \sum_s ( \Gamma_{s1} \notag\\
&& +\Gamma_{s2} ) [ \bar{f}_s(\varepsilon) + f_s(\varepsilon+U) ]  S_y  - \sum_s 2\sqrt{\Gamma_{s1}\Gamma_{s2}} \frac{1}{\pi} [ p_s(\varepsilon) \notag\\
&& + p_s(\varepsilon+U)] S_z , \label{eq:Sy_delta_t} 
\end{eqnarray}
\begin{eqnarray}
\dot{S}_z &=& \sum_s 2\sqrt{\Gamma_{s1}\Gamma_{s2}} \frac{1}{\pi} [p_s(\varepsilon) + p_s(\varepsilon+U) ] S_y - \sum_s (\Gamma_{s1} \notag\\
&&+ \Gamma_{s2}) [\bar{f}_s(\varepsilon) + f_s(\varepsilon+U)] S_z +\sum_s (\Gamma_{s1} - \Gamma_{s2} )  \{ f_s(\varepsilon)  \notag\\
&& 
\times P_0 -  [\bar{f}_s(\varepsilon) - f_s(\varepsilon+U)] \frac{P_1}{2} 
  -  \bar{f}_s(\varepsilon+U) P_2 \} , \label{eq:Sz_delta_t}
\end{eqnarray}
while for the populations, the equations become
\begin{eqnarray}
\dot{P}_0 &=&
\sum_s \{ ( \Gamma_{s1}  + \Gamma_{s2} ) \bar{f}_s(\varepsilon) P_1
-2 (\Gamma_{s1}  + \Gamma_{s2}) f_s(\varepsilon) P_0 \} \notag\\
&&+ \sum_s 4 \sqrt{\Gamma_{s1}\Gamma_{s2}} \bar{f}_s (\varepsilon) S_x 
+ \sum_s 2 ( \Gamma_{s1} - \Gamma_{s2} ) \bar{f}_s(\varepsilon) S_z , \notag\\
\end{eqnarray}
\begin{eqnarray}
\dot{P}_1 &=& 
\sum_s \{2 (\Gamma_{s1} +\Gamma_{s2}) f_s(\varepsilon) P_0
 -(\Gamma_{s1}+\Gamma_{s2}) [\bar{f}_s(\varepsilon)  \notag\\
 && +f_s(\varepsilon+U)] P_1
+2 (\Gamma_{s1} + \Gamma_{s2}) \bar{f}_s(\varepsilon+U) P_2 \} \notag\\
&&- \sum_s 4 \sqrt{\Gamma_{s1}\Gamma_{s2}} [\bar{f}_s(\varepsilon) - f_s(\varepsilon+U)] S_x \notag\\
&& - \sum_s 2 (\Gamma_{s1} -\Gamma_{s2} ) [\bar{f}_s(\varepsilon) - f_s(\varepsilon+U)] S_z , 
\end{eqnarray}
\begin{eqnarray}
\dot{P}_{2} &=&
\sum_s  \{ (\Gamma_{s1} + \Gamma_{s2}) f_s(\varepsilon+U) P_1  - 2 (\Gamma_{s1} + \Gamma_{s2} ) \notag\\
&&\times \bar{f}_s(\varepsilon+U) P_2 \} - \sum_s 4 \sqrt{\Gamma_{s1}\Gamma_{s2}} f_s(\varepsilon+U) S_x \notag\\
&& + \sum_s 2 ( \Gamma_{s2} - \Gamma_{s1}) f_s(\varepsilon+U) S_z .
\end{eqnarray}

Compared with the symmetric case [see Eqs.~(\ref{eq:Sx_symm})--(\ref{eq:P2_symm})], the small asymmetry in tunnelling rates, $\Gamma_{s1}-\Gamma_{s2} \sim \delta t$, couples $\{S_x,P_0,P_1\}$ to $\{S_y,S_z\}$. Specifically, $S_y$ and $S_x$ are coupled via principal values, $p_s(\varepsilon)$ and $p_s(\varepsilon+U)$, while $S_z$ is coupled to $P_0,P_1,P_2$ via Fermi-Dirac distribution functions. This means that the tunnelling asymmetry not only induces an additional effective magnetic field $B_z$ that causes spin rotations in the $xy$-plane, but also allows spin accumulation in the $z$ direction.


\begin{thebibliography}{9}
\bibitem{Stafford2007} C. A. Stafford, D. M. Cardamone, and S. Mazumdar, {\it The quantum interference effect transistor}, \href{http://iopscience.iop.org/article/10.1088/0957-4484/18/42/424014/pdf}{Nanotechnology {\bf 18}, 424014 (2007)}.

\bibitem{Vannucci15prb} L. Vannucci, F. Ronetti, G. Dolcetto, M. Carrega, and M. Sassetti, {\it Interference-induced thermoelectric switching and heat rectification in quantum Hall junctions}, \href{https://doi.org/10.1103/PhysRevB.92.075446}{Phys. Rev. B {\bf 92}, 075446 (2015)}.

\bibitem{SierraSanchez16prb} M. A. Sierra, M. Saiz-Bre\'{t}õn, F. Dom\'{i}nguez-Adame, and David S\'{a}nchez, {\it Interactions and thermoelectric effects in a parallel-coupled double quantum dot}, \href{https://doi.org/10.1103/PhysRevB.93.235452}{Phys. Rev. B {\bf 93}, 235452 (2016)}.

\bibitem{Lambert16Physique} C. J. Lambert, H. Sadeghi, Q. H. Al-Galiby, {\it Quantum-interference-enhanced thermoelectricity in single molecules and molecular films}, \href{https://doi.org/10.1016/j.crhy.2016.08.003}{C. R. Physique {\bf 17}, 1084 (2016)}.

\bibitem{Samuelsson17prl} P. Samuelsson, S. Kheradsoud, and B. Sothmann, {\it Optimal quantum interference thermoelectric heat engine with edge states}, \href{https://doi.org/10.1103/PhysRevLett.118.256801}{Phys. Rev. Lett. {\bf 118}, 256801 (2017)}.

\bibitem{Marcos18prb} A. Marcos-Vicioso, C. L\'{o}pez-Jurado, M. Ruiz-Garcia, and R. S\'{a}nchez, {\it Thermal rectification with interacting electronic channels: Exploiting degeneracy, quantum superpositions, and interference}, \href{https://doi.org/10.1103/PhysRevB.98.035414}{Phys. Rev. B {\bf 98}, 035414 (2018)}.

\bibitem{Doty09prl} M. F. Doty, J. I. Climente, M. Korkusinski, M. Scheibner, A. S. Bracker, P. Hawrylak, and D. Gammon, {\it Antibonding ground states in InAs quantum-dot molecules}, \href{https://doi.org/10.1103/PhysRevLett.102.047401}{Phys. Rev. Lett. {\bf 102}, 047401 (2009)}.

\bibitem{Guedon12nnano} C. M. Gu\'{e}don, H. Valkenier, T. Markussen, K. S. Thygesen, J. C. Hummelen, and S. J. Molen, {\it Observation of quantum interference in molecular charge transport}, \href{https://doi.org/10.1038/nnano.2012.37}{Nat. Nano. {\bf 7}, 305 (2012)}.

\bibitem{Garner2018} M. H. Garner, H. Li, Y. Chen, T. A. Su, Z. Shangguan, D. W. Paley, T. Liu, F. Ng, H. Li, S. Xiao, C. Nuckolls, L. Venkataraman, and G. C. Solomon {\it Comprehensive suppression of single-molecule conductance using destructive $\sigma$-interference}, \href{https://www.nature.com/articles/s41586-018-0197-9}{Nature {\bf 558}, 415 (2018)}. 

\bibitem{Miao2018} R. Miao, H. Xu, M. Skripnik, L. Cui, K. Wang, K. G. L. Pedersen, M. Leijnse, F. Pauly, K. W\"arnmark, E. Meyhofer, P. Reddy, and H. Linke, {\it Influence of Quantum Interference on the Thermoelectric Properties of Molecular Junctions}, \href{https://pubs.acs.org/doi/abs/10.1021/acs.nanolett.8b02207}{Nano Lett. {\bf 18}, 5666 (2018)}. 

\bibitem{Nilsson2010} H. A. Nilsson, O. Karlstr\"om, M. Larsson, P. Caroff, J. N. Pedersen, L. Samuelson, A. Wacker, L.-E. Wernersson, and H. Q. Xu, {\it Correlation-Induced Conductance Suppression at Level Degeneracy in a Quantum Dot}, \href{https://journals.aps.org/prl/pdf/10.1103/PhysRevLett.104.186804}{Phys. Rev. Lett. {\bf 104}, 186804 (2010)}.

\bibitem{Karlstrom2011} O. Karlstr\"om, J. N. Pedersen, P. Samuelsson, and A. Wacker, {\it Canyon of current suppression in an interacting two-level quantum dot}, \href{https://journals.aps.org/prb/abstract/10.1103/PhysRevB.83.205412}{Phys. Rev. B {\bf 83}, 205412 (2011)}.

\bibitem{BayerHawrylak01science} M. Bayer, P. Hawrylak, K. Hinzer, S. Fafard, M. Korkusinski, Z. R. Wasilewski, O. Stern, A. Forchel, {\it Coupling and entangling of quantum states in quantum dot molecules}, \href{https://doi.org/10.1126/science.291.5503.451}{Science {\bf 291}, 451 (2001)}.

\bibitem{Wiel02rmp} W. G. van der Wiel, S. D. Franceschi, J. M. Elzerman, T. Fujisawa, S. Tarucha, and L. P. Kouwenhoven, {\it Electron transport through double quantum dots}, \href{https://doi.org/10.1103/RevModPhys.75.1}{Rev. Mod. Phys. {\bf 75}, 1 (2003)}.

\bibitem{Cottet04prl} A. Cottet, W. Belzig, and C. Bruder, {\it Positive cross correlations in a three-terminal quantum dot with ferromagnetic contacts}, \href{https://doi.org/10.1103/PhysRevLett.92.206801}{Phys. Rev. Lett. {\bf 92}, 206801 (2004)}.

\bibitem{Cottet04epl} A. Cottet and W. Belzig, {\it Dynamical spin-blockade in a quantum dot with paramagnetic leads}, \href{https://doi.org/10.1209/epl/i2004-10009-9}{Europhys. Lett. {\bf 66}, 405 (2004)}.

\bibitem{Belzig05prb} W. Belzig, {\it Full counting statistics of super-Poissonian shot noise in multilevel quantum dots}, \href{https://doi.org/10.1103/PhysRevB.71.161301}{Phys. Rev. B {\bf 71}, 161301 (2005)}.

\bibitem{Li13srep} Z. Z. Li, C. H. Lam, T. Yu, and J. Q. You, {\it Detector-induced backaction on the counting statistics of a double quantum dot}, \href{https://doi.org/10.1038/srep03026}{Sci. Rep. {\bf 3}, 3026 (2013)}.

%\bibitem{Welack08prb} S. Welack, M. Esposito, U. Harbola, and S. Mukamel, {\it Interference effects in the counting statistics of electron transfers through a double quantum dot}, \href{https://doi.org/10.1103/PhysRevB.77.195315}{Phys. Rev. B {\bf 77}, 195315 (2008)}.

\bibitem{UrbanKonig09prb} D. Urban and J\"{u}rgen K\"{o}nig, {\it Tunable dynamical channel blockade in double-dot Aharonov-Bohm interferometers}, \href{https://doi.org/10.1103/PhysRevB.79.165319}{Phys. Rev. B {\bf 79}, 165319 (2009)}. 

\bibitem{Schaller09prb} G. Schaller, G. Kie\ss lich, and Tobias Brandes, {\it Transport statistics of interacting double dot systems: Coherent and non-Markovian effects}, \href{https://doi.org/10.1103/PhysRevB.80.245107}{Phys. Rev. B {\bf 80}, 245107 (2009)}. 

\bibitem{Burkard00prb} G. Burkard, D. Loss, and E. V. Sukhorukov, {\it Noise of entangled electrons: Bunching and antibunching}, \href{https://doi.org/10.1103/PhysRevB.61.R16303}{Phys. Rev. B {\bf 61}, R16303 (2000)}.

\bibitem{GaudreauHawrylak06prl} L. Gaudreau, S. A. Studenikin, A. S. Sachrajda, P. Zawadzki, A. Kam, J. Lapointe, M. Korkusinski, and P. Hawrylak, {\it Stability diagram of a few-electron triple dot}, \href{https://doi.org/10.1103/PhysRevLett.97.036807}{Phys. Rev. Lett. {\bf 97}, 036807 (2006)}.

\bibitem{ShimHawrylak09prb} Y. P. Shim, F. Delgado, and P. Hawrylak, {\it Tunneling spectroscopy of spin-selective Aharonov-Bohm oscillations in a lateral triple quantum dot molecule}, \href{https://doi.org/10.1103/PhysRevB.80.115305}{Phys. Rev. B {\bf 80}, 115305 (2009)}.

\bibitem{Michaelis06epl} B. Michaelis, C. Emary, and C. W. J. Beenakker, {\it All-electronic coherent population trapping in quantum dots}, \href{https://doi.org/10.1209/epl/i2005-10458-6}{Europhys. Lett. {\bf 73}, 677 (2006)}.

\bibitem{XuSham08nphys} X. Xu, B. Sun, P. R. Berman, D. G. Steel, A. S. Bracker, D. Gammon, and L. J. Sham, {\it Coherent population trapping of an electron spin in a single negatively charged quantum dot}, \href{https://doi.org/10.1038/nphys1054}{Nat. Phys. {\bf 4}, 692 (2008)}.

\bibitem{Bayer08nphys} M. Bayer, {\it Quantum optics with dots}, \href{https://doi.org/doi:10.1038/nphys1065}{Nat. Phys. {\bf 4}, 678 (2008)}.

\bibitem{Li11epl} Z. Z. Li, S. H. Ouyang, C. H. Lam, and J. Q. You, {\it Cooling a nanomechanical resonator by a triple quantum dot}, \href{https://doi.org/10.1209/0295-5075/95/40003}{Europhys. Lett. {\bf 95}, 40003 (2011)}.

\bibitem{SchultzOppen09prb}  M. G. Schultz and F. von Oppen, {\it Quantum transport through nanostructures in the singular-coupling limit}, \href{https://doi.org/10.1103/PhysRevB.80.033302}{Phys. Rev. B {\bf 80}, 033302 (2009)}.

\bibitem{Breuer02} H. P. Breuer and F. Petruccione, \textit{Theory of Open Quantum Systems} (Oxford, New York, 2002).

\bibitem{DarauDonariniGrifoni09prb} D. Darau, G. Begemann, A. Donarini, and M. Grifoni, {\it Interference effects on the transport characteristics of a benzene single-electron transistor}, \href{https://doi.org/10.1103/PhysRevB.79.235404}{Phys. Rev. B {\bf 79}, 235404 (2009)}.

\bibitem{BraunKonig04prb} M. Braun, J. K\"{o}nig, and J. Martinek, {\it Theory of transport through quantum-dot spin valves in the weak-coupling regime}, \href{https://doi.org/10.1103/PhysRevB.70.195345}{Phys. Rev. B {\bf 70}, 195345 (2004)}.

\bibitem{Hell15prb} M. Hell, B. Sothmann, M. Leijnse, M. R. Wegewijs, and J. K\"{o}nig, {\it Spin resonance without spin splitting}, \href{https://doi.org/10.1103/PhysRevB.91.195404}{Phys. Rev. B {\bf 91}, 195404} (2015).

\bibitem{Holubec18jltp} V. Holubec and T. Novotn\'{y}, {\it Effects of noise-induced coherence on the performance of quantum absorption refrigerators}, \href{https://doi.org/10.1007/s10909-018-1960-x}{J. Low Temp. Phys.  {\bf 192}, 147 (2018)}.

\bibitem{Blum96} K. Blum, {\it Density Matrix Theory and Applications} (Plenum Press, New York, 1996).

\bibitem{ScullyZubairy1997} M. O. Scully and M. S. Zubairy, {\it Quantum Optics} (Cambridge University Press, Cambridge, England, 1997).

\bibitem{qmeq17} G. Kir\v{s}anskas, J. N. Pedersen, O. Karlstr\"{o}m, M. Leijnse, and A. Wacker, {\it QmeQ 1.0: An open-source Python package for calculations of transport through quantum dot devices}, \href{https://doi.org/10.1016/j.cpc.2017.07.024}{Comput. Phys. Commun. {\bf 221}, 317 (2017)}.

\bibitem{Tokura07njp} Y. Tokura, H. Nakano, and T. Kubo, {\it Interference through quantum dots}, \href{https://doi.org/10.1088/1367-2630/9/5/113}{New J. Phys. {\bf 9}, 113 (2007)}.

\bibitem{Kubo06prb} T. Kubo, Y. Tokura, T. Hatano, and S. Tarucha, {\it Electron transport through Aharonov-Bohm interferometer with laterally coupled double quantum dots}, \href{https://doi.org/10.1103/PhysRevB.74.205310}{Phys. Rev. B {\bf 74}, 205310 (2006)}.

\bibitem{Leijnse2008} M. Leijnse and M. R. Wegewijs {\it Kinetic equations for transport through single-molecule transistors}, \href{https://journals.aps.org/prb/abstract/10.1103/PhysRevB.78.235424}{Phys. Rev. B {\bf 78}, 235424 (2008)}. 

\bibitem{Nazarov03} Y. V. Nazarov, {\it Quantum Noise in Mesoscopic Physics} (Kluwer, Dordrecht, 2003).

\bibitem{Chatterjee18prb} A. Chatterjee, S. N. Shevchenko, S. Barraud, R. M. Otxoa, F. Nori, J. J. L. Morton, and M. F. Gonzalez-Zalba, {\it A silicon-based single-electron interferometer coupled to a fermionic sea}, \href{https://doi.org/10.1103/PhysRevB.97.045405}{Phys. Rev. B {\bf 97}, 045405 (2018)}.
 
\end{thebibliography}
\end{document}